\documentclass
[superscriptaddress,secnumarabic,amssymb,amsmath,nobibnotes,aps,prd,showkeys,showpacs,nofootinbib,onecolumn,12pt]{revtex4}%
\usepackage{graphicx}
\usepackage{epsf}
\usepackage{bm}
\usepackage{amsmath}
\usepackage{amsfonts}
\usepackage{amssymb}%
\RequirePackage[colorlinks,citecolor=blue,urlcolor=blue,linkcolor=blue]{hyperref}
\providecommand{\U}[1]{\protect\rule{.1in}{.1in}}

\newcommand{\be}{\begin{equation}}
\newcommand{\ee}{\end{equation}}

\newcommand{\mincir}{\raise
-3.truept\hbox{\rlap{\hbox{$\sim$}}\raise4.truept\hbox{$<$}\ }}
\newcommand{\magcir}{\raise
-3.truept\hbox{\rlap{\hbox{$\sim$}}\raise4.truept\hbox{$>$}\ }}

\begin{document}
\title{Global dynamics and evolution for the Szekeres system with nonzero cosmological constant term}
\author{Andronikos Paliathanasis}
\email{anpaliat@phys.uoa.gr}
\affiliation{Institute of Systems Science, Durban University of Technology, Durban 4000,
South Africa }
\affiliation{Instituto de Ciencias F\'{\i}sicas y Matem\'{a}ticas, Universidad Austral de
Chile, Valdivia 5090000, Chile}
\author{Genly Leon}
\email{genly.leon@ucn.cl}
\affiliation{Departamento de Matem\'{a}ticas, Universidad Cat\'{o}lica del Norte, Avda.
Angamos 0610, Casilla 1280 Antofagasta, Chile.}

\begin{abstract}
The Szekeres system with cosmological constant term describes the evolution of the kinematic quantities for Einstein field equations in $\mathbb{R}^4$. In this study, we investigate the behavior of trajectories in the presence of cosmological constant. It has been shown that the Szekeres system is a Hamiltonian dynamical system. It admits at least two conservation laws, $h$ and $I_{0}$ which indicate the integrability of the Hamiltonian system. We solve the Hamilton-Jacobi equation, and we reduce the Szekeres system from $\mathbb{R}^4$ to an equivalent system defined in $\mathbb{R}^2$. Global dynamics are studied where we find that there exists an attractor in the finite regime only for positive valued cosmological constant and $I_0<2.08$. Otherwise, trajectories reach infinity. For $I_ {0}>0$ the origin of trajectories in $\mathbb{R}^2$ is also at infinity. Finally, we investigate the evolution of physical properties by using dimensionless variables different from that of Hubble-normalization conducing to a dynamical system in $\mathbb{R}^5$. We see that the attractor at the finite regime in $\mathbb{R}^5$ is related with the de Sitter universe for a positive cosmological constant.
\end{abstract}
\date{\today}
\maketitle

\section{Introduction}

\label{sec1}

A Szekeres system is a system of algebraic-differential
equations given by \cite{kras}
\begin{align}
\dot{\rho}_{D}+\theta\rho_{D} &  =0,\label{ss1.01}\\
\dot{E}+3E\sigma+\theta E+\frac{1}{2}\rho_{D}\sigma &  =0,\label{ss1.04}\\%
\dot{\theta}+\frac{\theta^{2}}{3}+6\sigma^{2}+\frac{1}{2}\left(  \rho
_{D}-2\Lambda\right)   &  =0,\label{ss1.02}\\
\dot{\sigma}-\sigma^{2}+\frac{2}{3}\theta\sigma+E &  =0,\label{ss1.03}
\end{align}
with algebraic constraint equation%
\begin{equation}
\frac{\theta^{2}}{3}-3\sigma^{2}+\frac{^{\left(  3\right)  }R}{2}-\Lambda
=\rho_{D}.\label{ss1.05}%
\end{equation}

Equations \eqref{ss1.01}- \eqref{ss1.05}  are the diagonal Einstein field
equations $G_{\mu\nu}-\Lambda g_{\mu\nu}=T_{\mu\nu}$ for a gravitational
model where the energy-momentum tensor $T_{\mu\nu}$ is that of a pressureless
inhomogeneous fluid, $\Lambda$ is the cosmological constant, and $G_{\mu\nu
}=R_{\mu\nu}-\frac{1}{2}Rg_{\mu\nu}$ is the Einstein tensor for the background
space%
\begin{equation}
ds^{2}=-dt^{2}+e^{2a(t,x,y,z)}dx^{2}+e^{2b(t,z,y,z)}(dy^{2}+dz^{2}%
).\label{ss1.06}%
\end{equation}
Scalars $a\left(  t,x,y,z\right)$ and $b\left(  t,x,y,z\right)$ satisfy
the field equations \eqref{ss1.01}- \eqref{ss1.05}, where $^{\left(  3\right)
}R$ is the spatial curvature for the three dimensional hypersurface of the
line element \eqref{ss1.06}, $\theta$ and $\sigma$ are the kinematic
quantities known as expansion rate and shear. For the comoving observer
$u^{\mu}=\delta_{t}^{\mu}$, $u^{\mu}u_{\mu}=-1$ they are defined as 
$\theta=\left(  \frac{\partial\alpha}{\partial t}\right)  +2\left(
\frac{\partial\beta}{\partial t}\right)$ and $\sigma^{2}=\frac{2}{3}\left(
\left(  \frac{\partial\alpha}{\partial t}\right)  -\left(  \frac{\partial
\beta}{\partial t}\right)  \right)  ^{2}\,$. Function $\rho_{D}=\rho\left(
t,x,y,z\right)$ describes the inhomogeneous energy density for the
pressureless fluid, and $E=E\left(  t,x,y,z\right)$ is the electric
component of the Weyl tensor~{$E_{\nu}^{\mu}=$ $Ee_{\nu}^{\mu}$.\ The dot in
equations }\eqref{ss1.01}- \eqref{ss1.04} remarks derivative with respect to
the time parameter. Equation \eqref{ss1.01} is the continuous equation for the
pressureless fluid, $T_{~;\nu}^{\mu\nu}=0$, while the rest of the equations
are the four diagonal Einstein field equation. Moreover, nondiagonal
components of the Einstein field equations are the propagation equations
\cite{lesame} {$h_{\mu}^{\nu}\sigma_{\nu;\alpha}^{\alpha}=\frac{2}{3}h_{\mu
}^{\nu}\theta_{;\nu}$, $~h_{\mu}^{\nu}E_{\nu;\alpha}^{\alpha}=\frac{1}%
{3}h_{\mu}^{\nu}\rho_{;\nu}$ where $h_{\mu\nu}$ is the projection tensor
defined as $h_{\mu\nu}=g_{\mu\nu}-u_{\mu}u_{\nu}$}.

The above gravitational model with the zero-valued cosmological term was
investigated by Szekeres in \cite{szek}. It was found that the spacetime
\eqref{ss1.06} describes inhomogeneous
Friedmann-Lema\^{\i}tre-Robertson-Walker (-like) universes  with one scale
factor which satisfies the Friedmann equations, or inhomogeneous anisotropic
Kantowski-Sachs (-like) with two scale factors. These spacetimes were one of
the first families of cosmological inhomogeneous exact solutions in the
literature. The case with a nonzero cosmological constant term investigated
\cite{sz2,sz2a} where it was found that because of the cosmological constant
inflationary solutions exist. The main characteristic, by the construction of the
Szekeres system is that their magnetic part of the Weyl tensor is zero, and
there is not any pressure term, which means that there is no information
propagation through gravitational waves or sound. Consequently, Szekeres
spacetimes belong to the family of Silent universes \cite{silent}. For other
extensions of the Szekeres system, we refer the reader in \cite{sz1,sz3,sz5}.
Inhomogeneous exact solutions are of special interest in cosmological studies (for a discussion we refer the reader to
\cite{rev1,dd6,dd7,dd8,dd9,per102,per103,per104}). 

Because of its importance in physical applications, the Szekeres system has
been widely investigated in the literature. With the use of Darboux
polynomials and Jacobi's last multiplier method, the integrability of the
Szekeres system with zero cosmological constant terms was found in
\cite{as1}. Specifically, three time-independent conservation laws where
derived which were used to reduce the four-dimensional Szekeres system into a
two-dimensional system. Moreover, the analytic solution for the
reduced two-dimensional system is presented in \cite{as2}. Furthermore, in
\cite{as3} a Lagrangian function was derived for the Szekeres system, which
means that the system \eqref{ss1.01}- \eqref{ss1.04} follows from a
variational principle. The corresponding conservation laws were derived
according to Noether's theorem. In addition, it was found that the
Szekeres system possesses the Painlev\'{e} property \cite{ls0}. Because the
Lagrangian function of the Szekeres system is a point-like Lagrangian, in
\cite{as4} it was used to quantize for the first time the Szekeres
system. The probability function was derived, while it was found that the
singular asymptotic solution is preferred by the quantization, on the other
hand, the quantum potentiality of the Szekeres system by Lie symmetries
\cite{ls1} was the subject of study in \cite{as5}. Moreover, the conservation
laws for the Szekeres system with a nonzero cosmological constant term were
derived recently in \cite{as6}.

In \cite{as7,as8}, Libre et al, performed a detailed analysis on the dynamics of
the Szekeres system with a zero cosmological constant in the Poincare disk. It
was found that the orbits come from the infinity of $\mathbb{R}^{4}$ and go to infinity. In this study, we extend this analysis by considering a nonzero cosmological constant term. As we shall see, for $\Lambda>0$, it is possible to have an attractor at the finite regime, which can be related to the de Sitter universe provided by the field equations.
The plan of the paper is as follows.

In Section \ref{sec2} we review previous results on the derivation of the
Lagrangian function which describes the dynamical system \eqref{ss1.01} and \eqref{ss1.04}. Moreover, we present the conservation laws, and we solve the
Hamilton-Jacobi equation to reduce the four-dimensional system into a
system of two first-order differential equations. The dynamics for the reduced
two-dimensional system are investigated in Sections \ref{sec3} and \ref{sec4}.
For the completeness of our analysis, we study the stationary points and the
evolution of the trajectories in the finite variables as also at the infinity
by using the variables of the Poincare disk. 

Furthermore, to understand the properties of the physical variables
during the evolution of the cosmological history, we define new dimensionless
variables, like that of the Hubble-normalization \cite{cop1}. In the new
variables, the dynamical system is defined in $\mathbb{R}^{5}$. Every stationary point on $\mathbb{R}^{5}$ corresponds to a specific epoch of the cosmological history, which is explicitly described by an analytic solution. From the stability properties of the stationary points, we can infer information about the evolution of the
model. Such analysis has been applied in various gravitational models in the past, with interesting results \cite{da1,da2,da3,da4,da5,da6,da7,da8,da9,da10}.
For the Szekeres system without the cosmological constant term, this analysis was
performed in \cite{silent}. However, in \cite{silent} the Hubble-normalization approach was applied, which means that in their study, the expansion rate has been limited to be only positive $\theta>0$, or negative $\theta<0$. However, given $\theta$ can change its sign, which means that static spacetimes with $\theta=0$ are possible, then we require new dimensionless variables \cite{da11} to avoid the blowing up of solutions as $\theta$ passes through zero. In Appendix \ref{app1} we present the
analysis for the case with $\Lambda=0$, and we recover previously
published results \cite{as7}. In  Section \ref{sec5a}, we present a detailed
analysis of the asymptotic solutions for field equations in dimensionless
variables different from that of the Hubble normalization. We find that
$\theta$ can change the sign in two stationary points which describe the Minkowski
universe and the closed Friedmann-Lema\^{\i}tre-Robertson-Walker (-like)
spacetime. Finally, in Section \ref{sec5}, we summarize our results and draw our conclusions.

\section{Hamiltonian system}

\label{sec2}

Solving equations \eqref{ss1.01} and \eqref{ss1.04}  for  $\theta$ and $\sigma$ (assuming $\rho_D  (6 E+\rho_D
   )\neq 0$) we obtain 
\begin{align}
    & \theta = -\frac{\dot{\rho_D}}{\rho_D}, \label{eq7}\\
    & \sigma = \frac{2 (E \dot{\rho_D} -\rho_D  \dot{E})}{\rho_D  (6 E+\rho_D
   )}. \label{eq8}
\end{align}
Substituting expressions \eqref{eq7} and \eqref{eq8} in equations \eqref{ss1.02} and \eqref{ss1.03}, we obtain 
the second order differential equations 
\begin{align}
 \ddot{\rho_D}= \rho   &  \left(\frac{24 (\rho 
   \dot{E}-E \dot{\rho_D} )^2}{\rho ^2 (6 E+\rho
   )^2}+\frac{1}{2} (\rho -2 \Lambda )+\frac{4 (\dot{\rho_D} )^2}{3 \rho
   ^2}\right), \label{ss1.02B}\\
\ddot{E}= \frac{1}{3 \rho  (6 E+\rho
   )^2}   & \Big[3 E \rho  (6 E+\rho
   )^2 (3 E-\Lambda +\rho ) \nonumber \\
   & +12 \rho  (12 E+\rho )
   (\dot{E})^2-4 E (9 E+\rho ) (\dot{\rho_D} )^2+8 \rho  (9
   E+\rho ) \dot{E} \dot{\rho_D} \Big]. \label{ss1.03B}
\end{align}
Using the parametrization (we assume $x\neq y, y\neq 0$)
\begin{equation}
 \rho_{D}\left(  x,y\right)  =\frac{6}%
{y^{2}\left(  y-x\right)  },~E\left(  x,y\right)  =-\frac{x}{y^{3}\left(
y-x\right)  },  
\end{equation}
where the inverse transformation is ~\cite{as3}
\begin{equation}
x\left(  \rho_{D},E\right)
=-\frac{6^{4/3}E}{\rho_{D}(6E+\rho_{D})^{1/3}}, ~y\left(  \rho_{D},E\right)
=\frac{6^{1/3}}{(6E+\rho_{D})^{1/3}},
\end{equation}
and replacing in equations \eqref{ss1.02B} and \eqref{ss1.03B}
we end with the system of two-second order differential equations \cite{as6}
\begin{align}
\ddot{x}-\frac{\Lambda}{3}x-\frac{2x}{y^{3}}  & =0,\label{s11a}\\
\ddot{y}-\frac{\Lambda}{3}y+\frac{1}{y^{2}}  & =0.\label{s12b}%
\end{align}

Equation \eqref{s12b} can be integrated as $I_{0}=\dot{y}^{2}-\frac{\Lambda
}{3}y^{2}-\frac{2}{y}$, which is a conservation law for the Szekeres system.
In addition, for the system \eqref{s11a}, \eqref{s12b} we can easily solve the
inverse problem and construct a Lagrangian function where under the variation
will produce the equations of motions. 

Indeed, the point-like Lagrangian which describes equations \eqref{s11a} and 
\eqref{s12b} is the two-dimensional Lagrangian%
\begin{equation}
L\left(  x,\dot{x},y,\dot{y}\right)  =\dot{x}\dot{y}+\frac{\Lambda}{3}%
xy-\frac{x}{y^{2}}.\label{reduced_lag}%
\end{equation}

Hence, we define the momentum $p_{x}=\frac{\partial L}{\partial\dot{x}}$  and
$p_{y}=\frac{\partial L}{\partial\dot{y}}$, i.e. $p_{x}=\dot{y}$ and
~$p_{y}=\dot{x}$, such that the Hamiltonian function $h=p_{x}\dot{x}+p_{y}%
\dot{y}-L$, to be%
\begin{equation}
h=\dot{x}\dot{y}-\frac{\Lambda}{3}xy+\frac{x}{y^{2}}.\label{ham1}%
\end{equation}
while the Hamilton's equations read%
\begin{equation}
\dot{x}=p_{y}~\ ,~\dot{y}=p_{x},
\end{equation}%
\begin{equation}
\dot{p}_{y}=\frac{\Lambda}{3}x+\frac{2x}{y}, \quad 
\dot{p}_{x}=\frac{\Lambda}{3}y-\frac{1}{y^{2}}.
\end{equation}
Because the Szekeres system is autonomous, the Hamiltonian function is a
conservation law which gives the energy of the system, that is, $\frac{dh}{dt}=0$. 

In the original variables the conservation laws become
\begin{equation}
I_{0}=\frac{2^{2/3}\left(  -3\Lambda-18\rho_{D}^{3}E-3\rho_{D}^{4}%
+24\theta\sigma+16\theta^{2}+9\sigma^{2}\right)  }{3\sqrt[3]{3}\rho_{D}%
^{2}(\rho_{D}+6E)^{2/3}}%
\end{equation}%
\begin{equation}
h=\frac{2^{2/3}\left(  3\rho_{D}\sigma(4\theta+3\sigma)+E\left(
6\Lambda-3\rho_{D}^{4}+42\theta\sigma-8\theta^{2}+36\sigma^{2}\right)
-18\rho_{D}^{3}E^{2}\right)  }{\sqrt[3]{3}\rho_{D}^{2}(\rho_{D}+6E)^{2/3}}.
\end{equation}
At this point we want to present another set of variables which are useful to
recognize the Hamiltonian formulation of the Szekeres system. We do the change
of variables $\rho=\exp\left(  \varrho\right), ~E=\frac{1}{6}\exp\left(
\varrho\right)  \left(  \exp\left(  -Z\right)  -1\right)$, and consider the
new variables $\varrho$ and $Z$. Therefore,  it follows $\theta=-\dot{\varrho}$ and $\sigma=\dot{Z}$.
\ Hence, the following Lagrangian function is a solution of the inverse
problem for the Szekeres system given by the Lagrange function%
\begin{equation}
L\left(  \varrho,\dot{\varrho},Z,\dot{Z}\right)  =e^{-Z-\frac{2}{3}\varrho
}\left(  3\dot{Z}-\dot{\varrho}\right)  \left(  6\dot{Z}+\dot{\varrho}\right)
-\frac{3}{2}e^{-4Z+\frac{1}{3}\varrho}\left(  3e^{3Z}-1\right).
\end{equation}
and its variation reproduces equations \eqref{ss1.01} and \eqref{ss1.04}.
Therefore, the four-dimensional Szekeres system is an integrable Hamiltonian
system.

\subsection{Hamilton-Jacobi equation}

 From \eqref{ham1} we write the time-independent Hamilton-Jacobi
equation%
\begin{equation}
\left(  \frac{\partial S}{\partial x}\right)  \left(  \frac{\partial
S}{\partial y}\right)  -\frac{\Lambda}{3}xy+\frac{x}{y^{2}}=h,
\end{equation}
while the conservation law $I_{0}$ reads%
\begin{equation}
\left(  \frac{\partial S}{\partial x}\right)  ^{2}-\frac{\Lambda}{3}%
y^{2}-\frac{2}{y}=I_{0}.%
\end{equation}
By integration we have 
\begin{align}
S\left(  x,y\right)    & =C_1(y) \pm \frac{x   \sqrt{3  {I}_0 y+\Lambda  y^3+6}}{\sqrt{3}\sqrt{y}},
\end{align}
where
\begin{equation}
\frac{ C_1'(y) \sqrt{3 I_0 y+\Lambda  y^3+6}}{\sqrt{3}
   \sqrt{y}}=\pm h.
\end{equation}
Therefore, the Action reads%
\begin{align}
S\left(  x,y\right)    & = \pm \int \frac{\sqrt{3}  h  \sqrt{y}}{\sqrt{3 I_0 y+\Lambda  y^3+6}} dy  \pm \frac{x   \sqrt{3  {I}_0 y+\Lambda  y^3+6}}{\sqrt{3}\sqrt{y}}. 
\end{align}
Consequently, the field equations are reduced to the following two-dimensional
system by using the Hamilton-Jacobi equations ~$\dot{x}=\frac{\partial
S}{\partial y}, ~\dot{y}=\frac{\partial S}{\partial x}$, that is,%
\begin{equation}
\dot{x}=\pm\frac{\sqrt{3}\left(  \Lambda xy^{3}-3x+3hy^{2}\right)  }%
{3y\sqrt{y\left(  6+3I_{0}y+\Lambda y^{3}\right)  }},
\end{equation}%
\begin{equation}
\dot{y}=\pm\frac{\sqrt{3}}{3y}\sqrt{y\left(  6+3I_{0}y+\Lambda y^{3}\right)  }.
\end{equation}
For the analysis in section \ref{sec3} we choose the branch $\epsilon=-1$.

\subsection{Special case:  vacuum ($\rho_D\equiv 0, E\neq 0$)}
For vacuum ($\rho_D\equiv 0$) equation \eqref{ss1.01} is trivially satisfied and defining $\Theta= 3 \sigma +\theta$ equations \eqref{ss1.04}, \eqref{ss1.02}, \eqref{ss1.03} and \eqref{ss1.05} become 
\begin{align}
\dot{E}+\Theta E &  =0, \label{eq28}\\
\dot{\Theta}+\frac{\Theta ^2}{3} -\Lambda +3 E&  =0, \label{eq29}\\
\dot{\sigma}+\frac{2 \Theta  \sigma }{3}-3 \sigma ^2+ E&  =0, \label{eq30}
\end{align}
with algebraic constraint equation%
\begin{equation}
\frac{\Theta ^2}{3}-2 \Theta  \sigma+\frac{^{\left(  3\right)  }R}{2}-\Lambda
=0. 
\end{equation}
Solving \eqref{eq29} for $\Theta$, defining $F$ such that $\sigma= -\dot{F}/F$, and defining 
\begin{align}
E= \frac{1}{Y^3}, \quad F=\sqrt[3]{\frac{X}{Y}},
\end{align}
the system \eqref{eq29}-\eqref{eq30} transforms to 
\begin{align}
  \ddot{X}-\frac{\Lambda}{3} X-\frac{2 X}{Y^3} &=0,\\
  \ddot{Y} -\frac{\Lambda}{3}  Y + \frac{1}{Y^2} &=0.
\end{align}
The previous analysis follows by setting $(x,y)=(X,Y)$. 

\subsection{Special case:  $(6 E+\rho_D)\equiv 0$}

In the special case $E=-\rho_D/6$ equations \eqref{ss1.01} and  \eqref{ss1.04} are the same. Equations \eqref{ss1.04}, \eqref{ss1.02} and \eqref{ss1.03} become 
\begin{align}
 \dot{\rho_D} + \theta  \rho_D  & =0 \label{eq35}\\
 \dot{\theta}+ \frac{\theta ^2}{3} +\frac{\rho_D}{2} +6 \sigma ^2 -\Lambda & =0, \label{eq36}\\
 \dot{\sigma}+ \frac{2 \theta \sigma }{3}-\frac{\rho_D}{6}-\sigma ^2 &=0, \label{eq37}
\end{align}
with algebraic constraint equation \eqref{ss1.05}.

Solving \eqref{eq35} for $\theta$, defining $F$ such that $\sigma= -\dot{F}/F$, and defining 
\begin{align}
\rho_D= -\frac{1}{2 Y (X+Y)^2}, \quad F=  \sqrt[3]{\frac{Y}{X+Y}},
\end{align}
the system \eqref{eq36}-\eqref{eq37} transforms to 
\begin{align}
\ddot{X}-\frac{\Lambda  X}{3}+\frac{1}{4 (X+Y)^2} &=0,\\
\ddot{Y}-\frac{\Lambda  Y}{3} -\frac{1}{4 (X+Y)^2}&=0.  
\end{align}
Defining 
\begin{equation}
    x= X+Y, \quad y=Y
\end{equation}
we obtain
\begin{align}
 \ddot{x}-\frac{\Lambda  x}{3} &=0, \label{s11aB}\\
\ddot{y}-\frac{\Lambda  y}{3} -\frac{1}{4x^2} &=0. \label{s12bB}
\end{align}
Equation \eqref{s11aB} can be integrated as $I_{0}=\dot{x}^{2}-\frac{\Lambda
}{3}x^{2}$.

The point-like Lagrangian which describes equations \eqref{s11aB} and 
\eqref{s12bB} is the two-dimensional Lagrangian%
\begin{equation}
L\left(x,\dot{x},y,\dot{y}\right)  =\dot{x}\dot{y}+\frac{\Lambda}{3}%
xy-\frac{1}{4 x}.\label{reduced_lagIII}%
\end{equation}

Hence we define the momentum $p_{x}=\frac{\partial L}{\partial\dot{x}}$  and
$p_{y}=\frac{\partial L}{\partial\dot{y}}$, i.e. $p_{x}=\dot{y}$ and
~$p_{y}=\dot{x}$, such that the Hamiltonian function $h=p_{x}\dot{x}+p_{y}%
\dot{y}-L$, to be%
\begin{equation}
h=\dot{x}\dot{y}-\frac{\Lambda}{3}xy+\frac{1}{4 x}.\label{ham1III}%
\end{equation}
while the Hamilton's equations read%
\begin{equation}
\dot{x}=p_{y}~\ ,~\dot{y}=p_{x},
\end{equation}%
\begin{equation}
\dot{p}_{y}=\frac{\Lambda}{3}x, \quad 
\dot{p}_{x}=\frac{\Lambda}{3}y + \frac{1}{4x^2}.
\end{equation}
Because the Szekeres system is autonomous, the Hamiltonian function is a
conservation law, the energy of the system, that is, $\frac{dh}{dt}=0$, i.e., $h$ is a constant. 

Therefore, the four-dimensional Szekeres system is an integrable Hamiltonian
system. From \eqref{ham1III} we write the time-independent Hamilton-Jacobi
equation%
\begin{equation}
\left(  \frac{\partial S}{\partial x}\right)  \left(  \frac{\partial
S}{\partial y}\right) -\frac{\Lambda}{3}xy+\frac{1}{4 x}=h.
\end{equation}
while the conservation law $I_{0}$ reads%
\begin{equation}
\left(  \frac{\partial S}{\partial y}\right)  ^{2}-\frac{\Lambda}{3}%
{x}^{2}=I_{0}.%
\end{equation}
Therefore, the action reads
\begin{align}
 & S(x,y)=    C_1(x) \pm  \frac{y   \sqrt{3 I_0+\Lambda  x^2}}{\sqrt{3}}, 
\\
&C_1(x)  = \pm \frac{1}{4}  \left(\frac{4 \sqrt{3}
   h \ln\left(\sqrt{\Lambda } \sqrt{3 I_0+\Lambda 
   x^2}+\Lambda  x\right)}{\sqrt{\Lambda }}+\frac{\ln\left(\sqrt{3}
   \sqrt{I_0} \sqrt{3 I_0+\Lambda  x^2}+3
   I_0\right)-\ln(x)}{\sqrt{I_0}}\right).
\end{align}
Consequently, the field equations are reduced to the following two-dimensional
system by using the Hamilton-Jacobi equations ~$\dot{x}=\frac{\partial
S}{\partial y}, ~\dot{y}=\frac{\partial S}{\partial x}$, that is,%
\begin{align}
\dot{x} & = \pm\frac{\sqrt{3 I_0+\Lambda  x^2}}{\sqrt{3}}, \\
    \dot{y} & =  \pm \frac{(12 h x+4 \Lambda  x y-3)}{4 \sqrt{9 I_0+3
   \Lambda  x^2}}. 
\end{align}
We choose the branch $\epsilon=-1$ and assume $3 I_0+\Lambda  x^2\geq 0$. 
That is, the physical cases are $\Lambda>0, \; x^2 \geq -\frac{3 I_0}{\Lambda}$ or $\Lambda<0, \; I_0>0, \; x^2 \leq -\frac{3 I_0}{\Lambda}$. In the second case, the variable $x$ is bounded. 

\section{Stability analysis at the finite regime}
\label{sec3}

\subsection{Case $(6 E+\rho_D)\neq 0$}

The reduced two-dimensional Szekeres system in the finite regime is described
by the following system of two first-order ordinary differential equations%
\begin{align}
\frac{dx}{d \bar{t}} &  =f\left(  x,y;h,I_{0}\right),  \label{sz.01}\\
\frac{dy}{d \bar{t}} &  =g\left(  x,y;h,I_{0}\right),  \label{sz.02}%
\end{align}
with%
\begin{align}
f\left(  x,y;h,I_{0}\right)   &  =-\sqrt{3}\left(  x\left(  y^{3}-3\right)
+3hy^{2}\right), \label{sz.03}\\
g\left(  x,y;h,I_{0}\right)   &  =-\sqrt{3}y\left(  \Lambda y^{3}%
+3I_{0}y+6\right), \label{sz.04}%
\end{align}
where we have selected the new independent variable $dt=\frac{\sqrt{y\left(
6+3I_{0}y+\Lambda y^{3}\right)  }}{3y}d\bar{t}$.

The stationary points $P=\left(  x\left(  P\right), y\left(  P\right)
\right)$ for the dynamical system \eqref{sz.01}, \eqref{sz.02} are given by
the roots of the algebraic equations $f\left(  x,y,h,I_{0}\right)  =0$ and
$g\left(  x,y;h,I_{0}\right)  =0$. Hence, for a nonzero cosmological constant
$\Lambda$ and nonzero value for the conservation law $I_{0},~$the stationary
points are derived%
\[
P_{0}=\left(  0,0\right),
\]%
\[
P_{1}=\left(  \frac{3hy\left(  P_{1}\right)  ^{2}}{\Lambda y\left(
P_{1}\right)  ^{3}-3},\left(  \sqrt{\Lambda^{3}\left(  I_{0}^{3}%
+9\Lambda\right)  }-3\Lambda^{2}\right)  ^{\frac{1}{3}}\left(  \frac
{1}{\Lambda}-\frac{I_{0}}{\left(  \sqrt{\Lambda^{3}\left(  I_{0}^{3}%
+9\Lambda\right)  }-3\Lambda^{2}\right)  ^{\frac{2}{3}}}\right)  \right),
\]%
\[
P_{2}=\left(  \frac{3hy\left(  P_{2}\right)  ^{2}}{\Lambda y\left(
P_{2}\right)  ^{3}-3},\frac{I_{0}\left(  \Lambda+i\sqrt{3}\Lambda\right)
+\left(  -1+i\sqrt{3}\right)  \left(  \sqrt{\Lambda^{3}\left(  I_{0}%
^{3}+9\Lambda\right)  }-3\Lambda^{2}\right)  ^{\frac{2}{3}}}{2\Lambda\left(
\sqrt{\Lambda^{3}\left(  I_{0}^{3}+9\Lambda\right)  }-3\Lambda^{2}\right)
^{\frac{1}{3}}}\right),
\]
and%
\[
P_{3}=\left(  \frac{3hy\left(  P_{3}\right)  ^{2}}{\Lambda y\left(
P_{3}\right)  ^{3}-3},\frac{I_{0}\left(  \Lambda-i\sqrt{3}\Lambda\right)
-\left(  1+\sqrt{3}\right)  \left(  \sqrt{\Lambda^{3}\left(  I_{0}%
^{3}+9\Lambda\right)  }-3\Lambda^{2}\right)  ^{\frac{2}{3}}}{2\Lambda\left(
\sqrt{\Lambda^{3}\left(  I_{0}^{3}+9\Lambda\right)  }-3\Lambda^{2}\right)
^{\frac{1}{3}}}\right).
\]
Stationary point $P_{0}$ is always real. However, that is not always true for
points $P_{1}$, $P_{2}$ and $P_{3}$. Thus for various values of the free
variables $\left\{  h, I_{0}\right\}$, the number of stationary points can be two
or four. We proceed by assuming that the cosmological constant is positive,
say $\Lambda=1,$ or negative, e.g., $\Lambda=-1$.

\subsubsection{Positive $\Lambda$}

Let us assume now that $\Lambda=1$. Then, the stationary points read%
\[
P_{0}=\left(  0,0\right), 
\]%
\[
P_{1}=\left(  \frac{3hy\left(  P_{1}\right)  ^{2}}{\Lambda y\left(
P_{1}\right)  ^{3}-3},\left(  \sqrt{\left(  I_{0}^{3}+9\right)  }%
-3^{2}\right)  ^{\frac{1}{3}}\left(  1-\frac{I_{0}}{\left(  \sqrt{\left(
I_{0}^{3}+9\right)  }-3^{2}\right)  ^{\frac{2}{3}}}\right)  \right), 
\]%
\[
P_{2}=\left(  \frac{3hy\left(  P_{2}\right)  ^{2}}{\Lambda y\left(
P_{2}\right)  ^{3}-3},\frac{I_{0}\left(  1+i\sqrt{3}\right)  +\left(
-1+i\sqrt{3}\right)  \left(  \sqrt{\left(  I_{0}^{3}+9\right)  }-3^{2}\right)
^{\frac{2}{3}}}{2\left(  \sqrt{\left(  I_{0}^{3}+9\Lambda\right)  }%
-3^{2}\right)  ^{\frac{1}{3}}}\right), 
\]
and%
\[
P_{3}=\left(  \frac{3hy\left(  P_{3}\right)  ^{2}}{\Lambda y\left(
P_{3}\right)  ^{3}-3},\frac{I_{0}\left(  1-i\sqrt{3}\right)  -\left(
1+\sqrt{3}\right)  \left(  \sqrt{\left(  I_{0}^{3}+9\right)  }-3\right)
^{\frac{2}{3}}}{2\left(  \sqrt{\left(  I_{0}^{3}+9\Lambda\right)  }%
-3^{2}\right)  ^{\frac{1}{3}}}\right)  ~.
\]
Consequently, for $I_{0}$ we find  three intervals $\mathcal{A}=\left(
-\infty,-2.08\right)$,~$\mathcal{B}=\left(  -2.08,0\right)$ and
$\mathcal{C}=\left(  0,+\infty\right)$. Thus, in the interval $\mathcal{A}$,
points $P_{1},~P_{2},~P_{3}$ are real, that is $\left\{  P_{1},P_{2}%
,P_{3}\right\}  \in\mathbb{R}$. For $I_{0}\in\mathcal{B}$, $P_{2}\in\mathbb{R}$ while $\left\{  P_{1},P_{3}\right\}  \in
\mathbb{C}$ with $\operatorname{\Im}\left(  P_{1}\right)  \operatorname{\Im}\left(
P_{3}\right)  \neq0$, were we use the operator $\operatorname{\Im}(z)$ for the imaginary part of $z\in\mathbb{C}$. Finally, for $I_{0}\in\mathcal{C}$ the only real point
is $P_{1}$, while $\left\{  P_{2},P_{3}\right\}  \in\mathbb{C}.$ So, for values of the integration constant in the interval $\mathcal{A}$,
equations \eqref{sz.01} and \eqref{sz.02} admit four real stationary points
$\left\{  P_{0},P_{1},P_{2},P_3\right\}$ while for $I_{0}\in\mathcal{B}$ or
$I_{0}\in\mathcal{C}$, the stationary points are two, that is, the set of
points $\left\{  P\,_{0},P_{2}\right\}$ and $\left\{  P_{0},P_{1}\right\}$,  respectively.

\begin{figure}[ptb]
\centering\includegraphics[width=1\textwidth]{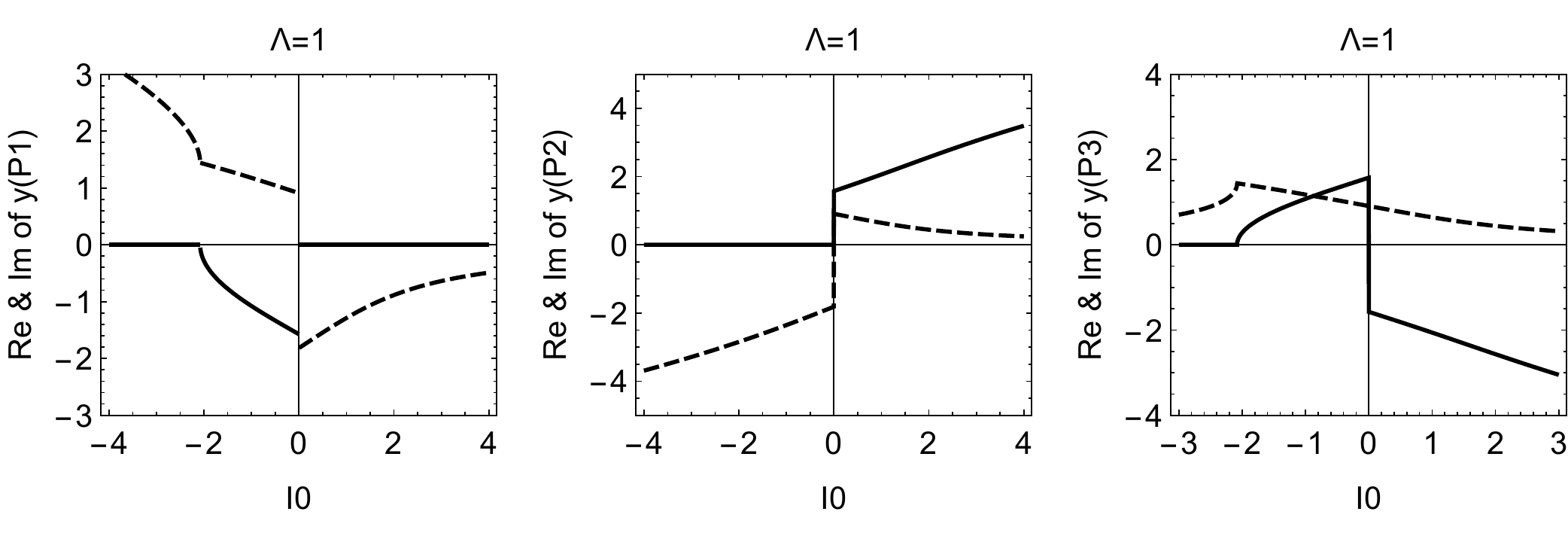}\caption{Imaginary (solid lines) and real parts (dashed lines) for the variables
$y\left(  P_{1}\right)$, $y\left(  P_{2}\right)$ and $y\left(
P_{3}\right)$ are represented. We observe that points $\left\{  P_{1},P_{2},P_{3}\right\}
\in\mathbb{R}$ for $I_{0}<-2.08$. Moreover, when $-2.08<I_{0}<0$, only $P_{2}$
is real, while for $I_{0}>0$, $P_{1}$ $\ $is a real point. By definition
$y>0$, hence, points $P_{1}$ and $P_{3}$ are physically accepted points for
$I_{0}<-2.08$. Finally, for $I_{0}>-2.08$ no stationary points in the finite
regime are physically accepted.}%
\label{fig1}%
\end{figure}

In Fig. \ref{fig1} we present the imaginary and real parts for the variables
$y\left(  P_{1}\right)$,~$y\left(  P_{2}\right)$ and $y\left(
P_{3}\right)$. Because $y$ should be positive, then the
only physically accepted  points in the finite regime are $P_{0}$ for arbitrary
$I_{0}$ and $P_{1}$, $P_{3}$ for $I_{0}<-2.08$.

\begin{figure}[ptb]
\centering\includegraphics[width=1\textwidth]{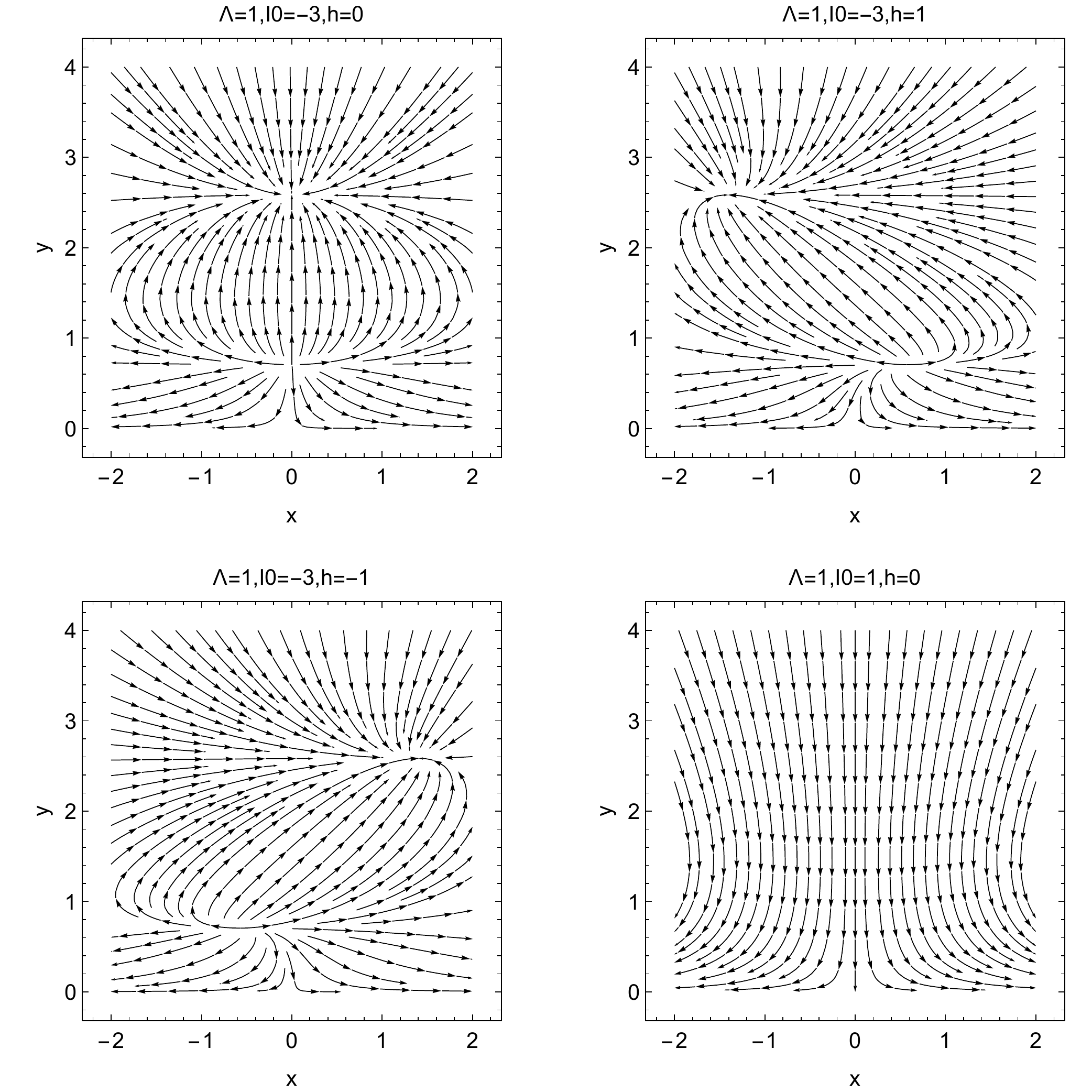}\caption{Phase-space
portraits for the dynamical system \eqref{sz.01}, \eqref{sz.02} in the $x-y$
plane, for different values of the free parameters $I_{0}$ and $h$ and
positive cosmological constant, $\Lambda=1$. For $I_{0}=-3$ we observe that
the unique attractor is point $P_{1}$ while for $I_{0}=1$ there is not any
attractor in the finite regime, and the unique stationary point is the saddle
point $P_{0}$. }%
\label{fig2A}%
\end{figure}

To understand the stability properties of the stationary points, we
linearize the system \eqref{sz.01}, \eqref{sz.02} around $P+\delta
P$. We derive the eigenvalues of the two-dimensional matrix, namely
$e_{1}\left(  P\right)$ and $e_{2}\left(  P\right)$, that we use to
investigate the stability of the point.

For point $P_{0}$ the eigenvalues are $e_{1}\left(  P_{0}\right)  =-6\sqrt{3}$
and $e_{2}\left(  P_{0}\right)  =3\sqrt{3}$ where we infer that the $P_{0}$ is
a saddle point. For the stationary point $P_{1}$ the eigenvalues of the
linearized system are~$e_{1}\left(  P_{1}\right)  =3\sqrt{3}e\left(
I_{0}\right)$ and $e_{1}\left(  P_{1}\right)  =6\sqrt{3}e\left(
I_{0}\right)$ with $e\left(  I_{0}\right)  =\left(  3+\frac{I_{0}\left(
\left(  \sqrt{\left(  I_{0}^{3}+9\right)  }-3\right)  ^{\frac{2}{3}}%
-I_{0}\right)  }{\left(  \sqrt{\left(  I_{0}^{3}+9\right)  }-3\right)
^{\frac{1}{3}}}\right)$. Thus, for $I_{0}<-2.08$  we derive
$\operatorname{\Re}\left(  e\left(  I_{0}\right)  \right)  <0$  from where we
infer that $P_{1}$ is an attractor. We use the operator $\operatorname{\Re}(z)$ for the real part of $z\in\mathbb{C}$. Finally, for point $P_{3}$ we find that
the two eigenvalues have always positive real part for $I_{0}<-2.08,$ which
means that $P_{3}$ is always a source.

In Fig. \ref{fig2A} we present the phase-space portraits for the dynamical
system \eqref{sz.01}, \eqref{sz.02}  in the finite regime  for various values
of the free variables $I_{0}$ and $h$. We observe that for $I_{0}<-2.08$,
point $P_{1}$ is the unique attractor while for $I_{0}>-2.08$ there is not any
attractor in the finite regime and the unique stationary point is the saddle
point $P_{0}$. The value of parameter $h$ changes only the location of the
stationary point on the $x-$direction and does not change the stability of the
stationary points.

\subsubsection{Negative $\Lambda$}

\begin{figure}[ptb]
\centering\includegraphics[width=1\textwidth]{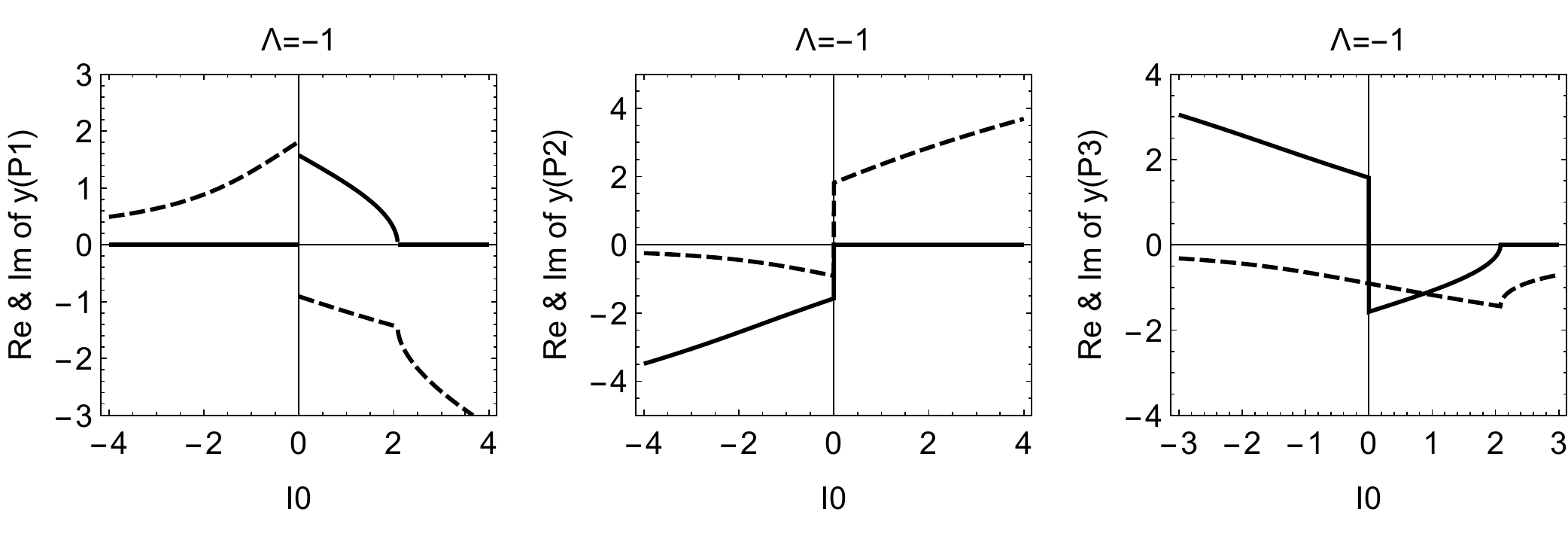}\caption{Imaginary (solid lines) and real parts (dashed lines) for the variables
$y\left(  P_{1}\right)$, $y\left(  P_{2}\right)$ and $y\left(
P_{3}\right)$ are represented. We observe that points $\left\{  P_{1},P_{2},P_{3}\right\}
\in\mathbb{R}$ for $I_{0}>-2.08$. Moreover, when $0<I_{0}<2.08$, only $P_{2}$
is real, while for $I_{0}<0$, $P_{1}$ is a real point. Due to $y>0$,
hence, points $P_{1}$ and $P_{2}$ are physically accepted points for $I_{0}<0$
and $I_{0}>0$, respectively.}%
\label{fig3}%
\end{figure}

In Fig. \ref{fig3} we present the real and imaginary parts of the three
stationary points $P_{1},~P_{2}$ and $P_{3}$ for $\Lambda=-1.$ We have 
three intervals $\mathcal{A}_{\left(  -1\right)  }\mathcal{=}\left(
-\infty,0\right)$,~$\mathcal{B}_{\left(  -1\right)  }\mathcal{=}\left(
0,2.08\right)$ and $\mathcal{C}_{\left(  -1\right)  }=\left(  2.08,+\infty
\right)$ for the values of $I_{0}$. Thus, for $I_{0}\in\mathcal{A}_{\left(
-1\right)  }$, $\ P_{1}\in\mathbb{R}$ and $\left\{  P_{2},P_{3}\right\}  \in
\mathbb{C}$. When $I_{0}\in\mathcal{B}_{\left(  -1\right)  }$, $P_{2}\in \mathbb{R}$ and $\left\{  P_{1}, ~P_{3}\right\}  \in\mathbb{C}$. Finally, when $I_{0}\in\mathcal{C}_{\left(  -1\right)  }$, $\left\{
P_{1},P_{2},P_{3}\right\}  \in \mathbb{R}$. However, because $y$ is positive defined, it follows that for
$I_{0}\in\mathcal{A}_{\left(  -1\right)  }$, the stationary points are two,
$\left\{  P_{0},P_{1}\right\}$ while for $I_{0}\in\mathcal{B}_{\left(
-1\right)  }\cup\mathcal{C}_{\left(  -1\right)  }$ there are two physical
accepted stationary points, namely \ $\left\{  P_{0},P_{2}\right\}$.

\begin{figure}[ptb]
\centering\includegraphics[width=1\textwidth]{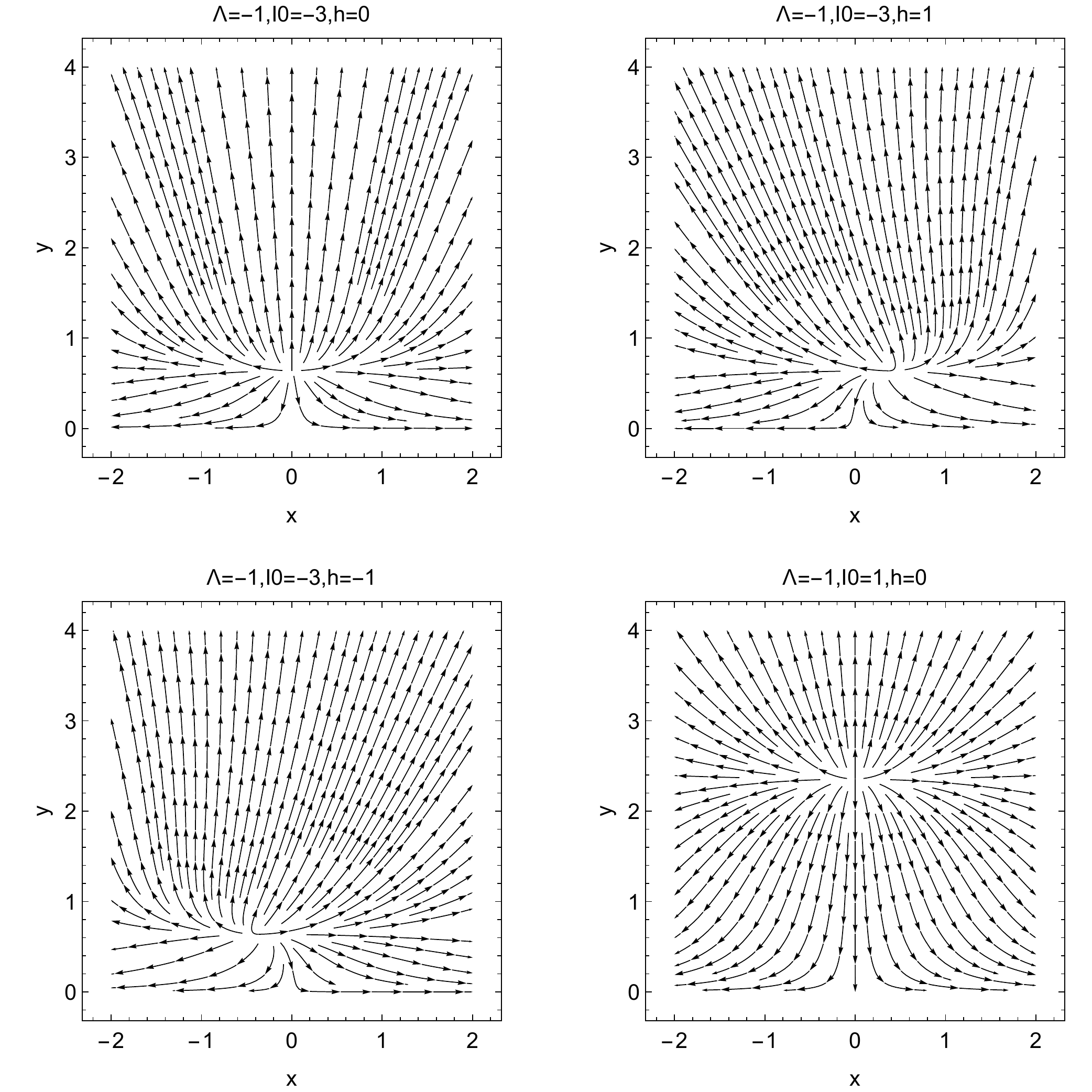}\caption{Phase-space
portraits for the dynamical system \eqref{sz.01} and  \eqref{sz.02} in the $x-y$
plane, for different values of the free parameters $I_{0}$ and $h$ and
negative cosmological constant, $\Lambda=-1$. For $I_{0}=-3$ and $I_{0}=1$ we
observe that $P_{1}$ and $P_{2}$ points are sources, respectively, while
$P_{0}$ is a saddle point. Thus, attractors do not exist in the finite regime
for negative $\Lambda$. }%
\label{fig4}%
\end{figure}

We investigate the stability properties for the stationary points, and we find
that $P_{0}$ is always a saddle point, while the stationary point $P_{1}$ for
$I_{0}<0$, and $P_{2},~$for $I_{0}>0$, have positive eigenvalues when they are
physically accepted. Consequently,  attractors do not exist in the finite regime for a negative cosmological constant. Phase-space portraits of
equations \eqref{sz.01}, \eqref{sz.02} \ for the variables $\left(
x,y\right)$ and for $\Lambda=-1$ are presented in Fig. \ref{fig4}, from
where we observe no attractor exists in the finite regime.

We repeat the calculations for $I_{0}=0$ and $\Lambda=1$. We find that there
is not any physically accepted stationary point except the saddle point
$P_{0}$, while for negative cosmological constant, i.e. $\Lambda=-1$, there
exist two physically accepted  stationary points $\left\{P_{0},P_{1}\right\}
$. Point $P_{0}$ is a saddle point, while $P_{1}$ is a source.

\subsection{Stability analysis at the finite regime $(6E+\rho_{D})=0$}

\label{sec3NEW}

The reduced two-dimensional Szekeres system in the finite regime is described
by the following system of two first-order ordinary differential equations%
\begin{align}
\frac{dx}{d\bar{t}} &  =f\left(  x,y;h,I_{0}\right)  ,\label{sz.01NEW}\\
\frac{dy}{d\bar{t}} &  =g\left(  x,y;h,I_{0}\right)  ,\label{sz.02NEW}%
\end{align}
with%
\begin{align}
f\left(  x,y;h,I_{0}\right)   &  =-3I_{0}-\Lambda x^{2}\label{sz.03NEW}\\
g\left(  x,y;h,I_{0}\right)   &  =-3hx-\Lambda xy+\frac{3}{4},\label{sz.04NEW}%
\end{align}
where we have selected the new independent variable $dt=\sqrt{3}\sqrt
{3I_{0}+\Lambda x^{2}}d\bar{t}$. \ However, by definition in order $\rho
_{D}>0$, it follows that $y<0$. 

The stationary points $Q=\left(  x\left(  Q\right)  ,y\left(  Q\right)
\right)  ~$for the dynamical system (\ref{sz.01NEW}), (\ref{sz.02NEW}) at the
finite regime are%
\[
Q_{\pm}=\left(  \pm\sqrt{-\frac{3I_{0}}{\Lambda}},-\frac{3h}{\Lambda}\pm
\frac{1}{4}\sqrt{-\frac{3}{I_{0}\Lambda}}\right)  .~
\]
From hypothesis $3 I_0+\Lambda  x^2\geq 0$ we have the physical cases $\Lambda>0, \; x^2 \geq -\frac{3 I_0}{\Lambda}$ or $\Lambda<0, \; I_0>0, \; x^2 \leq -\frac{3 I_0}{\Lambda}$. In the first case $Q_\pm$  exist for $I_0<0$ and the allowed region is $x^2 \geq -\frac{3 I_0}{\Lambda}, y<0$.  In the second case, the variable $x$ is bounded and $Q_\pm$ always exist and the allowed region is  $x^2 \leq -\frac{3 I_0}{\Lambda}, y<0$. Easily from the linearized system around the stationary points we derive the
eigenvalues $e_{1}\left(  Q_{\pm}\right)  =\pm \sqrt{-3\Lambda I_{0}}$ and
$e_{1}\left(  Q_{\pm}\right)  =\pm 2\sqrt{-3\Lambda I_{0}}$, which means that
point $Q_{-}$ is always a source and $Q_+$ is a sink.

When $I_0\geq 0, \Lambda\geq 0$ there are no stationary points at the finite region. 

The choice $I_0<0, \Lambda< 0$ give differential equations with coefficients in $\mathbb{C}$. Therefore, the choice of these parameters is forbidden.  

\section{Compactification}
\label{sec4}

\subsection{Poincare disk:  $(6 E+\rho_D)\neq 0$}

To investigate the dynamics at the infinity regime, we define the new
compactified variables $x=\frac{\rho}{\sqrt{1-\rho^{2}}}\cos\Phi$  and $y=\frac{\rho}{\sqrt{1-\rho^{2}}}\sin\Phi$, in which $\Phi\in\lbrack0,\pi]$
and $\rho\in\left[  0,1\right].$  As 
$x,y$ take infinite  values we have $\rho\rightarrow1$. In the set of variables $\left\{  \rho
,\Phi\right\}$ the field equations \eqref{sz.01} and \eqref{sz.02} read%
\begin{small}
\begin{equation}
\rho^{\prime}=\frac{\sqrt{3}}{2}\rho\left(  2\rho\cos^{2}\Phi\left(  \cos
\Phi\left(  \left(  3I_{0}-\Lambda\right)  \rho^{2}-3I_{0}\right)  -2h\sin
\Phi\left(  1-\rho^{2}\right)  \right)  -3\left(  1-\rho^{2}\right)
^{\frac{3}{2}}\left(  1+3\cos\left(  2\Phi\right)  \right)  \right)
,\label{sz.05}%
\end{equation}%
\end{small}
\begin{equation}
\Phi^{\prime}=3\sqrt{3}\cos\Phi\left(  \cos\Phi\left(  I_{0}\sin\Phi-h\cos
\Phi\right)  \rho+3\sin\Phi\sqrt{1-\rho^{2}}\right), \label{sz.06}%
\end{equation}
where $\Phi^{\prime}=\frac{d\Phi}{d\tau}$,~~$dt=\sqrt{1-\rho^{2}}d\tau$.

For values of $\rho$ near to one, equations \eqref{sz.05} and \eqref{sz.06}
become%
\begin{align}
&\rho^{\prime}=-\sqrt{3}\Lambda\cos^{3}\Phi~\equiv F\left(  \Phi\right)
,\label{sz.07}%
\\
& \Phi^{\prime}=3\sqrt{3}\cos^{2}\Phi\left(  I_{0}\sin\Phi-h\cos\Phi\right)
\equiv G\left(  \Phi\right)  .\label{sz.08}%
\end{align}
We continue our analysis by investigating the stationary points of equation
\eqref{sz.08}. From \eqref{sz.07} we know that when $F\left(\Phi\right) >0$, that is, $\rho^{\prime}\geq0$, the trajectories will still be 
at infinity, while for $F\left(  \Phi\right)<0$, that is, $\rho^{\prime
}<0$, this means that $\rho$ decreases, that is we move far from the infinite
regime. We observe that for $\Lambda<0$, $F\left(  \Phi\right)  >0$,  that is,
the infinity regime attracts the trajectories. It is an interesting result
because as we found before for negative cosmological constant there are not 
attractors at the finite regime.

The stationary points of equation \eqref{sz.08} are $\Phi=\Phi_{0}$ such that $G\left(  \Phi_{0}\right)  =0$, that is, $\Phi_{0}^{1}=\frac{\pi}{2}$
and $\Phi_{0}^{2}=\arctan\frac{h}{I_{0}}$ for $I_{0}\neq0$. Thus, by
definition $\frac{h}{I_{0}}\geq0$. In the special case $h=0$, it
can be found a third stationary point $\Phi_{0}^{3}=\pi$. Moreover, we calculate
$F\left(  \Phi_{0}^{1}\right)  =0$, while $F\left(  \Phi_{0}^{2}\right)
=-\sqrt{3}\Lambda\left(  1+\left(  \frac{h}{I_{0}}\right)  ^{2}\right)  ^{-3}%
$. 
\begin{figure}[ptb]
\centering\includegraphics[width=1\textwidth]{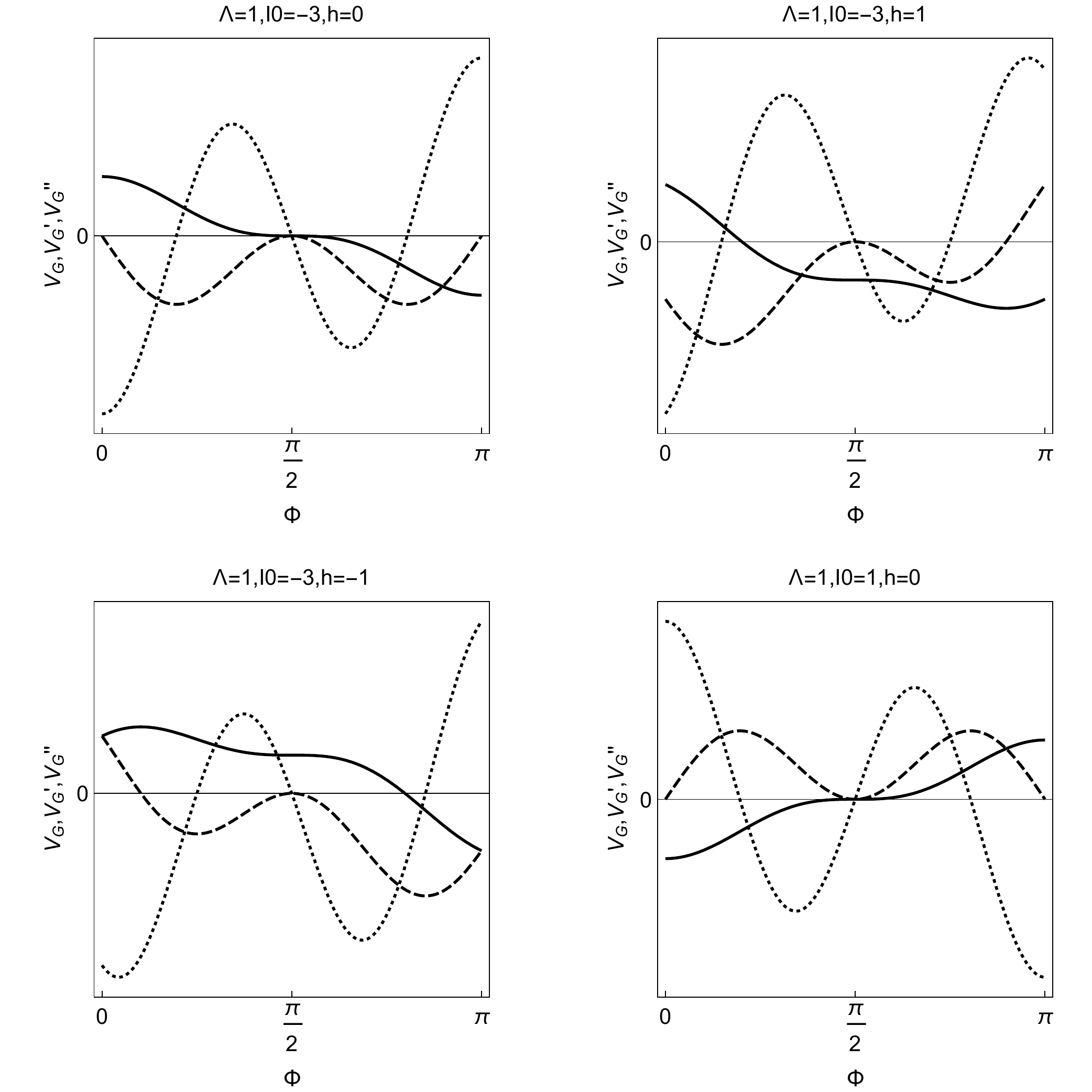}\caption{Qualitative
evolution of potential function $V_{G}\left(  \Phi\right)$ and  its
derivatives $\frac{dV_{G}}{d\Phi}$ and $\frac{d^{2}V_{G}}{d\Phi^{2}}$ in the interval $\left[  0,\pi\right]  .$}%
\label{fig7}%
\end{figure}

We continue with the study of \eqref{sz.08}.

We define the potential function $V_{G}\left(  \Phi\right)  =\int G\left(
\Phi\right)  d\Phi$, that is,
\begin{equation}
V_{G}\left(  \Phi\right)  =-\frac{\sqrt{3}}{4}\left(  4I_{0}\cos^{2}%
\Phi+h\left(  9\sin\Phi+\sin3\Phi\right)  \right).\label{sz.09}%
\end{equation}
For $\Phi_{0}^{1}$ we derive $V_{G}\left(  \Phi_{0}^{1}\right)  =-2\sqrt{3}h$,
which can be easily checked that it is not a minimum of the potential.
Furthermore,  $\frac{dG\left(  \Phi\right)  }{d\Phi}$  changes sign around 
$\Phi_{0}^{1}$, which means that point $\Phi_{0}^{1}$ is a saddle point.
However, because $\frac{dG\left(  \Phi\right)  }{d\Phi}|_{\Phi\rightarrow
\Phi_{0}^{1}}=0$, we can easily define  a stable manifold. In Fig. \ref{fig7}
we present the qualitative analysis for the potential $V_{G}$ and its two
derivatives  for specific values of the free variables $\left\{
I_{0},h\right\}$. We find that for $I_{0}<0$, $\Phi_{0}^{1}$ is minimum of 
the potential $V_{G}\left(\Phi\right)$ in the interval $\left[0,\frac{\pi
}{2}\right]$. While, for $I_{0}>0$, $\Phi_{0}^{1}$ is minimum of  the
potential $V_{G}\left(  \Phi\right)$ in the interval $\left[\frac{\pi}%
{2},0\right]$. Thus, for $I_{0}<0,$ $\Phi_{0}^{1}$ is an attractor in
$\left[0,\frac{\pi}{2}\right], $ i.e. $x>0$. While, for $I_{0}>0$, $\Phi
_{0}^{1}$ is an attractor for initial conditions in the interval $\Phi
\in\left[  \frac{\pi}{2},0\right]$, i.e. $x<0.$

As far as point $\Phi_{0}^{2}$ is concerned, we observe that $\frac{dG\left(
\Phi\right)  }{d\Phi}|_{\Phi\rightarrow\Phi_{0}^{2}}=\frac{3\sqrt{3}I_{0}%
}{\sqrt{1+\left(  \frac{h}{I_{0}}\right)  ^{2}}}$, which means that the point
is physically accepted  and it is an attractor in the surface with $\rho=1$, when
$I_{0}<0,$~$h<0$. On the other hand, for $h=0$, it follows that $\Phi_{0}^{2}$
is an attractor when $I_{0}>0$, while for $I_{0}<0$ the attractor is $h<0$. 

\begin{figure}[ptb]
\centering\includegraphics[width=1\textwidth]{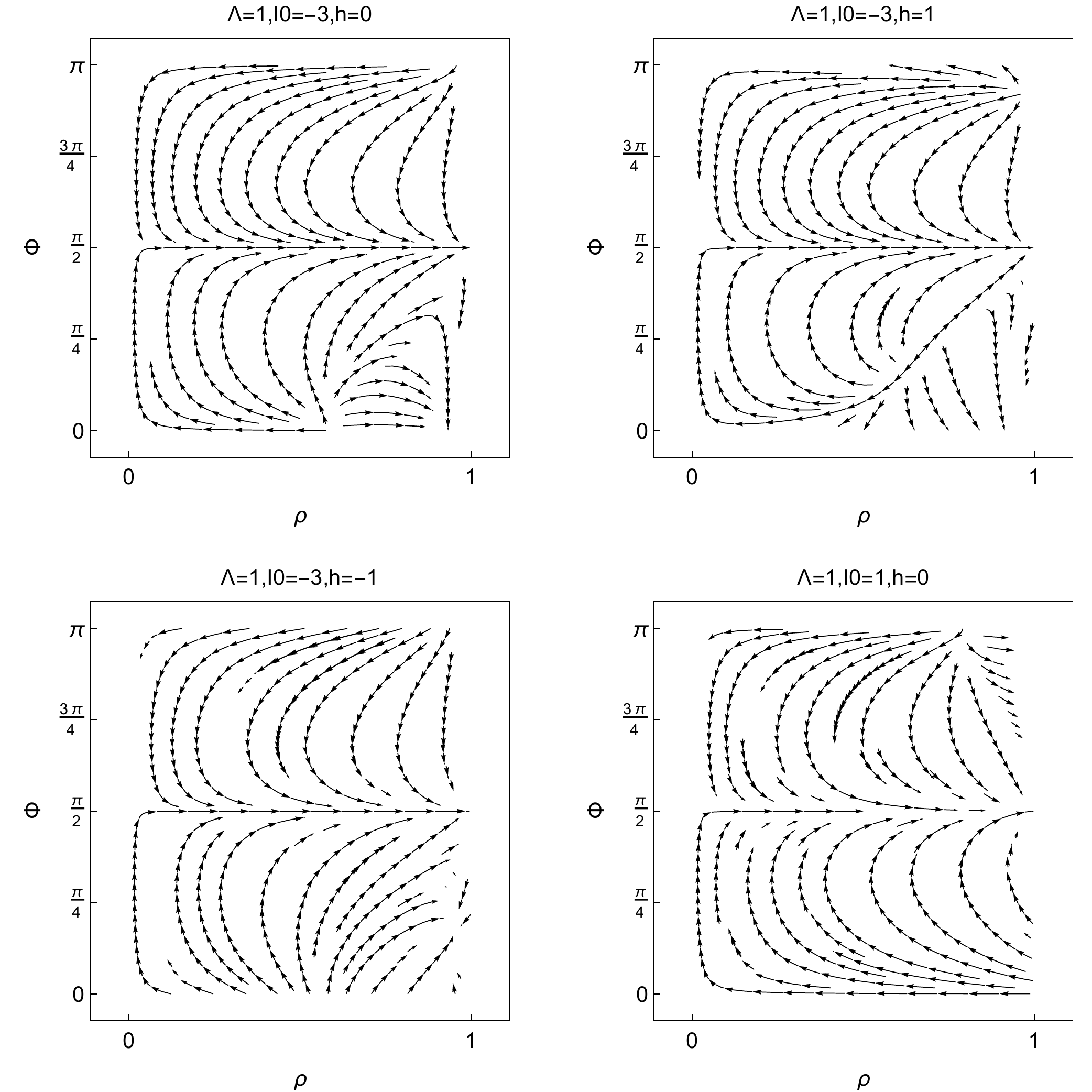}\caption{Phase-space
portraits for the dynamical system \eqref{sz.05}, \eqref{sz.06} in the
compactified variables $\left(  \rho,\Phi\right)$ for different values of
the free parameters $I_{0}$ and $h$ and positive cosmological constant,
$\Lambda=1$. }%
\label{fig5}%
\end{figure}

\begin{figure}[ptb]
\centering\includegraphics[width=1\textwidth]{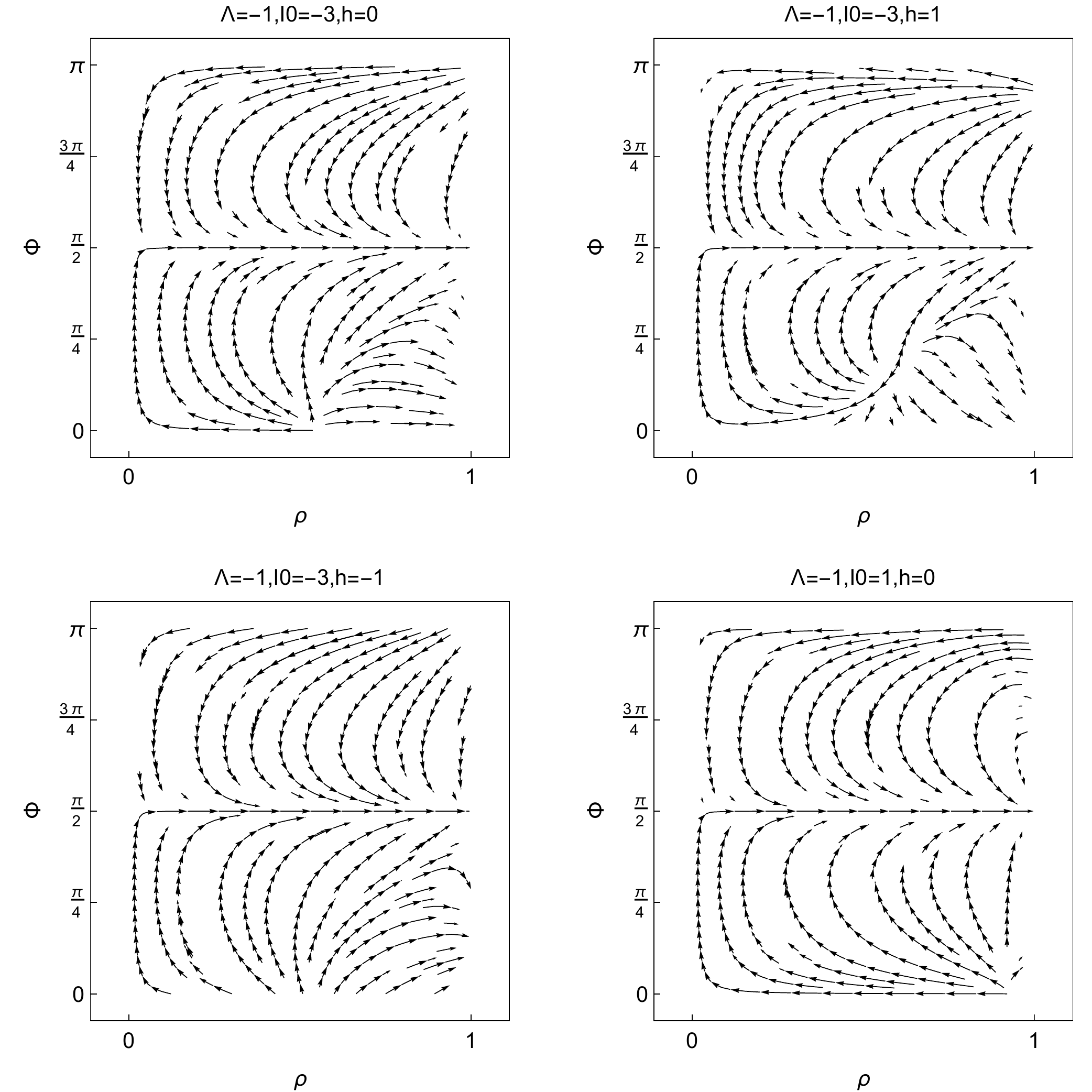}\caption{Phase-space
portraits for the dynamical system \eqref{sz.05}, \eqref{sz.06} in the
compactified variables $\left(  \rho,\Phi\right)$ for different values of
the free parameters $I_{0}$ and $h$ and negative cosmological constant,
$\Lambda=-1$. }%
\label{fig6}%
\end{figure}

In Figs. \ref{fig5} and \ref{fig6} we present the phase-space diagrams of the
field equations in the compactified variables $\left(  \rho,\Phi\right)$ for
various values of the free variables. The main results are summarized in the
following proposition.

\textit{Proposition}:  For Szekeres system \eqref{sz.01},
\eqref{sz.02} with a positive cosmological constant, that is, $\Lambda=1$, the trajectories can be originated at the finite or infinite regime with attractor point at the finite regime given by $P_{1}$ if $I_{0}<-2.08$  When $I_{0}>-2.08$, the
trajectories reach the infinity regime. On the contrary, for negative
cosmological constant, $\Lambda=-1$, the trajectories may start at the finite
or infinity regime and end at the infinity regime.

\begin{figure}[ptb]
\centering\includegraphics[width=1\textwidth]{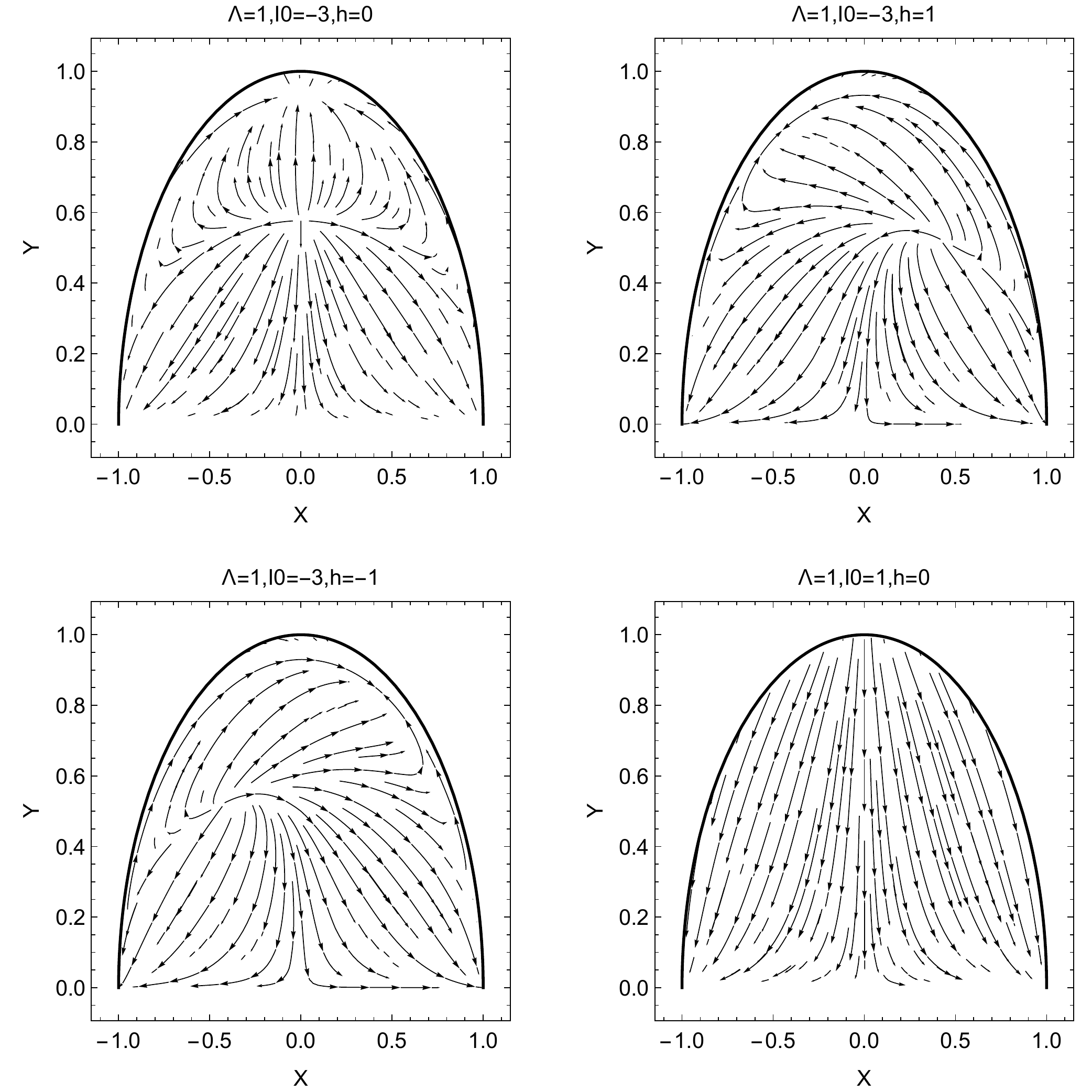}\caption{Phase-space
portraits for Szekeres system associated to \eqref{sz.01} and \eqref{sz.02} in the compactified variables $\left(
X,Y\right)$ for different values of the free parameters $I_{0}$ and $h$ and
positive cosmological constant, $\Lambda=1$. }%
\label{fig8}%
\end{figure}

\begin{figure}[ptb]
\centering\includegraphics[width=1\textwidth]{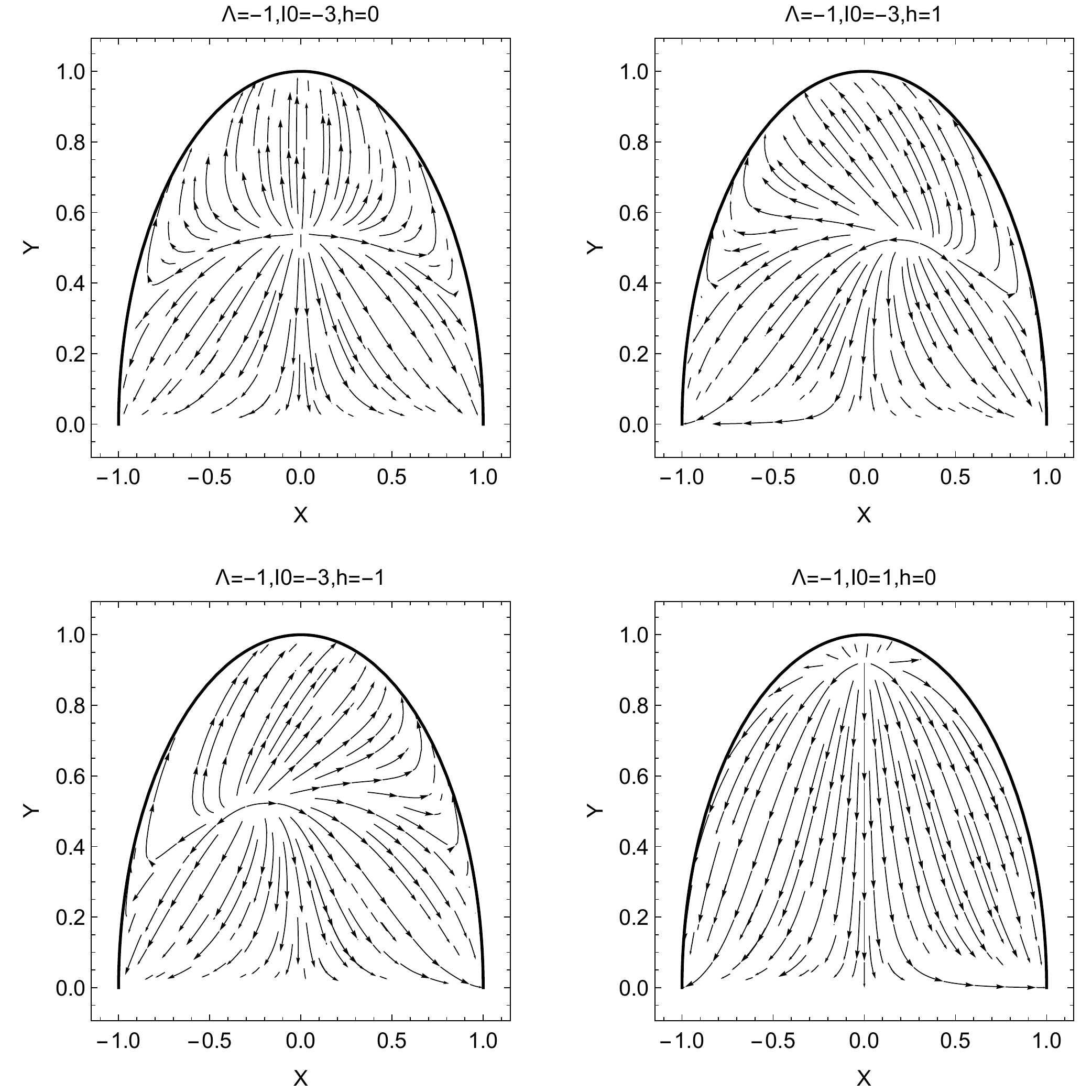}\caption{Phase-space
portraits for Szekeres system  associated to \eqref{sz.01} and \eqref{sz.02} in the compactified variables $\left(
X,Y\right)$ for different values of the free parameters $I_{0}$ and $h$ and
negative cosmological constant, $\Lambda=-1$. }%
\label{fig9}%
\end{figure}

In order to understand better the evolution of the trajectories, in Figs.
\ref{fig8} and \ref{fig9} we present the phase space portrait in the
compactified variables $\left(  X,Y\right)$ which are defined as $x=\frac
{X}{\sqrt{1-X^{2}-Y^{2}}}$ and $y=\frac{Y}{\sqrt{1-X^{2}-Y^{2}}}$ in which
$\left\vert X\right\vert \leq1$ and $0\leq Y\leq1$.

\subsection{Poincare disk:  $(6 E+\rho_D)= 0$}

The reduced two-dimensional Szekeres system in the finite regime is described
by the system of two first-order ordinary differential equations \eqref{sz.01NEW} and \eqref{sz.02NEW} with $f\left(  x,y;h,I_{0}\right)$ defined by \eqref{sz.03NEW} and $g\left(  x,y;h,I_{0}\right)$ defined by \eqref{sz.04NEW}
where we have selected the new independent variable $dt=\sqrt{3}\sqrt
{3I_{0}+\Lambda x^{2}}d\bar{t}$. 

Recall, 
 \begin{enumerate}
     \item From hypothesis $3 I_0+\Lambda  x^2\geq 0$ we have the physical cases: 
     \begin{enumerate}
         \item $\Lambda>0, \; x^2 \geq -\frac{3 I_0}{\Lambda},\; y<0$ 
         \item $\Lambda<0, \; I_0>0, \; x^2 \leq -\frac{3 I_0}{\Lambda},\; y<0$.
     \end{enumerate}
     
     \item When $I_0\geq 0, \Lambda\geq 0$ there are no stationary points at the finite region. 
     
     \item The choice $I_0<0, \Lambda< 0$ give differential equations with coefficients in $\mathbb{C}$. Therefore, the choice of these parameters is forbidden. 
 \end{enumerate}

\subsubsection{Case a.1: $\Lambda>0,\; I_0>0,\; y<0$}

In this case, the phase-plane  is the lower half-plane 
\begin{equation}
  H_-:=  \{(x,y)\in \mathbb{R}^2, x\in \mathbb{R}, y\leq 0 \},
\end{equation}
where we have attached the boundary set $y=0$. There are not stationary points in $H_-$. 
To investigate the dynamics at the infinity regime, we define the new
compactified variables $x=\frac{\rho}{\sqrt{1-\rho^{2}}}\cos\Phi$  and $y=\frac{\rho}{\sqrt{1-\rho^{2}}}\sin\Phi$, in which $\Phi\in\lbrack -\pi,\pi]$
and $\rho\in\left[  0,1\right].$ Since the physical region is the lower half plane $y\leq 0$, we have the physical interval $\Phi\in\lbrack -\pi,0]$. As 
$x,y$ take infinite  values we have $\rho\rightarrow1$. In the set of variables $\left\{  \rho
,\Phi\right\}$ the field equations \eqref{sz.01NEW} and \eqref{sz.02NEW} with $f\left(  x,y;h,I_{0}\right)$ defined by \eqref{sz.03NEW} and $g\left(  x,y;h,I_{0}\right)$ defined by \eqref{sz.04NEW} become 
\begin{align}
 \frac{d\rho}{d \bar{t}}& =  3 h \rho  \left(\rho ^2-1\right) \sin (\Phi ) \cos (\Phi
   ) \nonumber \\
   & +\frac{1}{4} \sqrt{1-\rho ^2} \left(-4 \cos (\Phi ) \left(\rho ^2
   (\Lambda -3 I_0)+3 I_0\right)-3 \left(\rho ^2-1\right)
   \sin (\Phi )\right), \label{69}\\
 \frac{d\Phi}{d \bar{t}}& =-3 h \cos ^2(\Phi ) + \frac{3 \sqrt{1-\rho ^2} (4 I_0
   \sin (\Phi )+\cos (\Phi ))}{4 \rho }. \label{70}
\end{align}
The leading terms as $\rho\rightarrow 1 ^-$ are
\begin{align}
& \frac{d\rho}{d \bar{t}}=   - \sqrt{2} \Lambda  \cos (\Phi ) \sqrt{1-\rho} - 6  h \cos (\Phi )
   \sin (\Phi )  (1-\rho)+O\left((1-\rho)^{3/2}\right),\\
& \frac{d\Phi}{d \bar{t}}= -3 h \cos ^2(\Phi ) + \frac{3 \sqrt{1-\rho ^2} (4 I_0
   \sin (\Phi )+\cos (\Phi ))}{4 \rho }+O\left((1-\rho)^{3/2}\right).
\end{align}
The points at infinity satisfy $\cos (\Phi
   )=0$ for $h\neq 0$. 

In terms of the Poincaré variables 
\begin{equation}
    X= \frac{x}{\sqrt{1+x^2 + y^2}}, \quad Y=\frac{y}{\sqrt{1+x^2 + y^2}}, \quad X\in[-1,1], \quad Y\in[-1,1]\end{equation}
we obtain 
\begin{align}
  &  \frac{d X}{d \bar{t}}=3 h X^2 Y+\sqrt{1-X^2-Y^2} \left(3 I_0 \left(X^2-1\right)-\frac{1}{4} X (4 \Lambda  X+3 Y)\right), \label{74}\\
   &  \frac{d Y}{d \bar{t}}= - 3 h X \left(1-Y^2\right)+\frac{1}{4} \sqrt{1-X^2-Y^2} (Y (12 I_0 X-4 \Lambda  X-3 Y)+3). \label{75} 
\end{align}
The physical region is $X^2+Y^2\leq 1, Y\leq 0$. 

\begin{figure}[ptb]
\centering\includegraphics[width=1\textwidth]{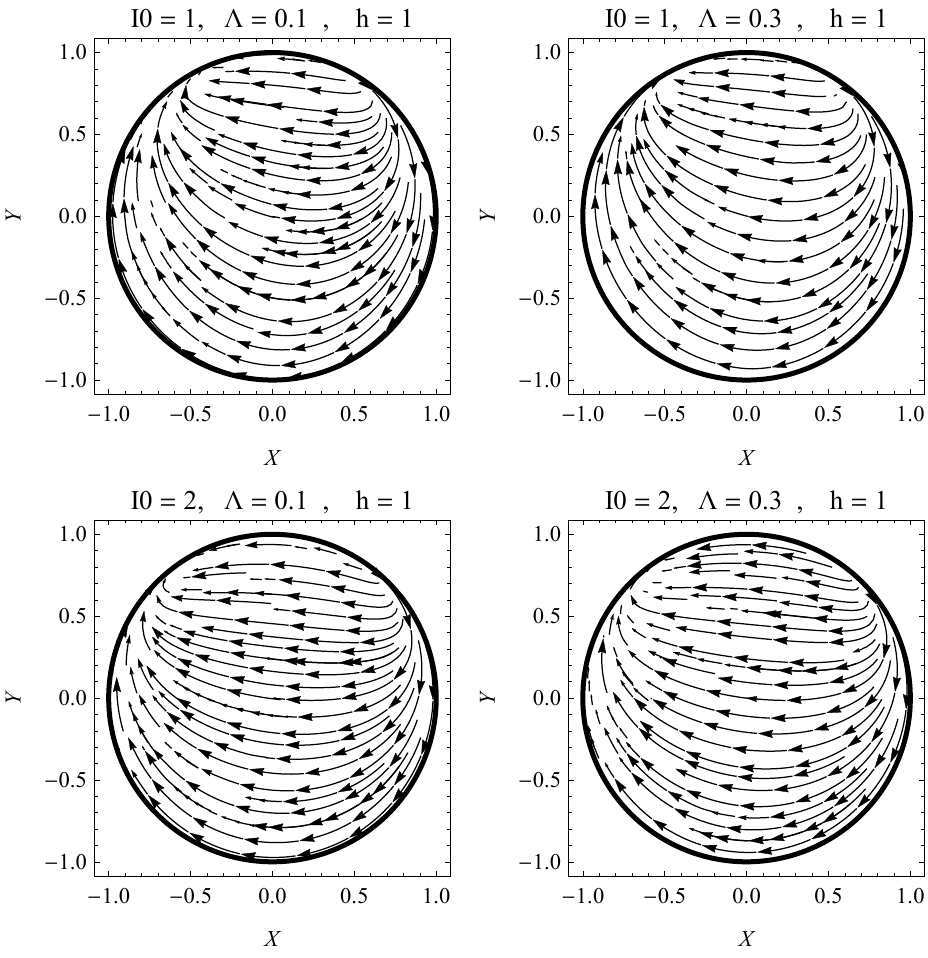}\caption{Phase-space
portraits for Szekeres system associated to \eqref{74} and \eqref{75} in the compactified variables $\left(
X,Y\right)$ for positive values of the free parameters $I_{0}$ and $\Lambda$ and $h=1$ corresponding to \emph{Case a.1}. }%
\label{fig8b}%
\end{figure}

In Fig.  \ref{fig8b} the phase-space portraits for Szekeres system associated to \eqref{74} and \eqref{75} is presented in the compactified variables $\left(
X,Y\right)$ for positive values of the free parameters $I_{0}$ and $\Lambda$ and $h=1$ corresponding to \emph{Case a.1}. 

\begin{figure}[ptb]
\centering\includegraphics[width=1\textwidth]{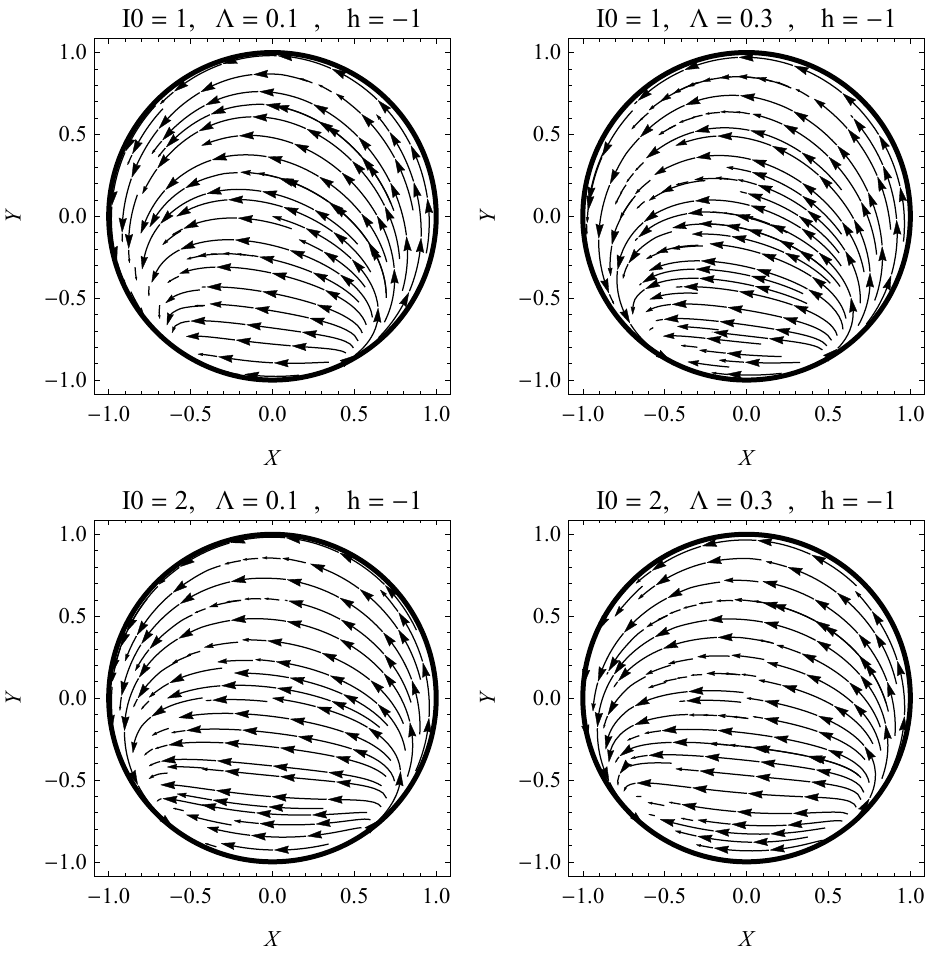}\caption{Phase-space
portraits for Szekeres system associated to \eqref{74} and \eqref{75} in the compactified variables $\left(
X,Y\right)$ for positive values of the free parameters $I_{0}$ and $\Lambda$ and $h=-1$ corresponding to \emph{Case a.1}. }%
\label{fig9b}%
\end{figure}

In Fig.   \ref{fig9b} the phase-space 
portraits for Szekeres system associated to \eqref{74} and \eqref{75} is presented in the compactified variables $\left(
X,Y\right)$ for positive values of the free parameters $I_{0}$ and $\Lambda$ and $h=-1$ corresponding to \emph{Case a.1}.

\subsubsection{Case a.2: $\Lambda>0,\; I_0\leq 0, \; x^2 \geq -\frac{3 I_0}{\Lambda}, \; y<0$}

The dynamics is given also by \eqref{69}-\eqref{70} in terms of $(\rho, \Phi)$ or by  \eqref{74}-\eqref{75} in terms of $(X,Y)$. The points at infinity satisfy $\cos (\Phi
   )=0$ for $h\neq 0$, but also we have the stationary points at the finite region $Q_{\pm}$.

\begin{figure}[ptb]
\centering\includegraphics[width=1\textwidth]{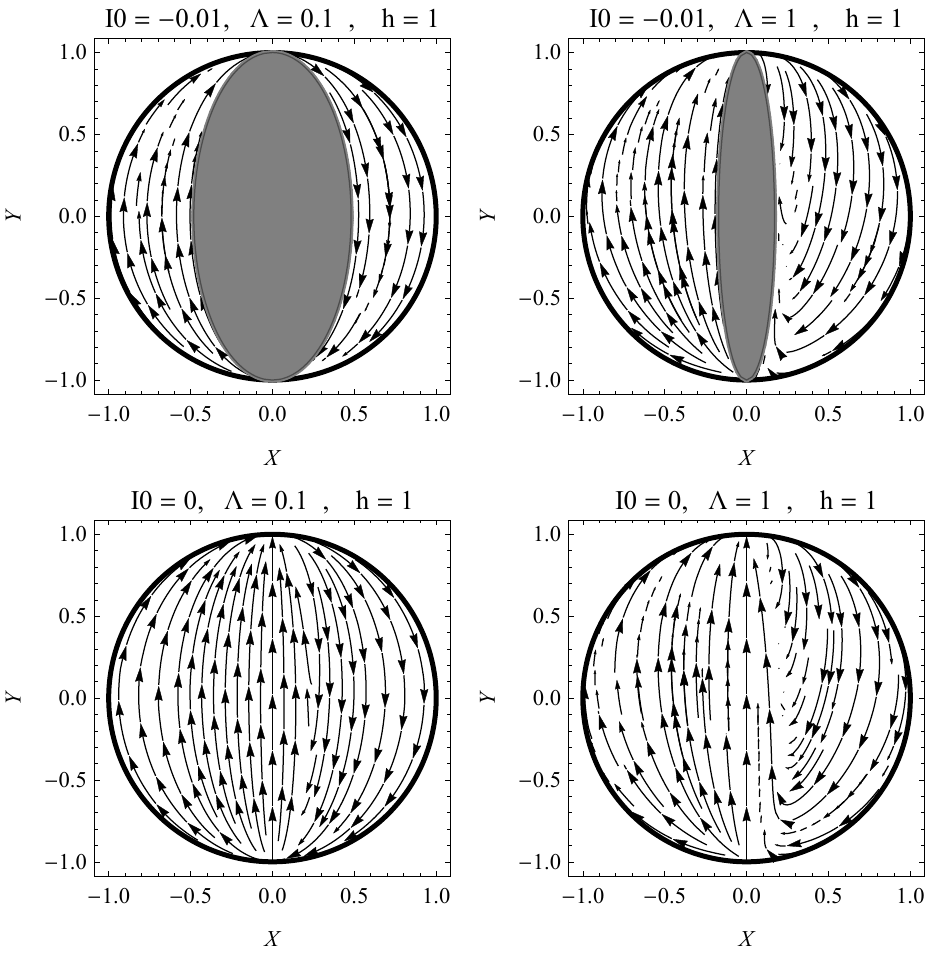}\caption{Phase-space
portraits for Szekeres system associated to \eqref{74} and \eqref{75} in the compactified variables $\left(
X,Y\right)$ for $I_{0}\leq 0$ and $\Lambda>0$ and $h=1$ corresponding to \emph{Case a.2}.  The gray region is the forbidden  region $x^2 \leq -\frac{3 I_0}{\Lambda}$.}%
\label{fig8c}%
\end{figure}

In Fig.  \ref{fig8c} the phase-space
portraits for Szekeres system associated to \eqref{74} and \eqref{75} is showed in the compactified variables $\left(
X,Y\right)$ for $I_{0}\leq 0$ and $\Lambda>0$ and $h=1$ corresponding to \emph{Case a.2}.  The gray region is the forbidden  region $x^2 \leq -\frac{3 I_0}{\Lambda}$.

\begin{figure}[ptb]
\centering\includegraphics[width=1\textwidth]{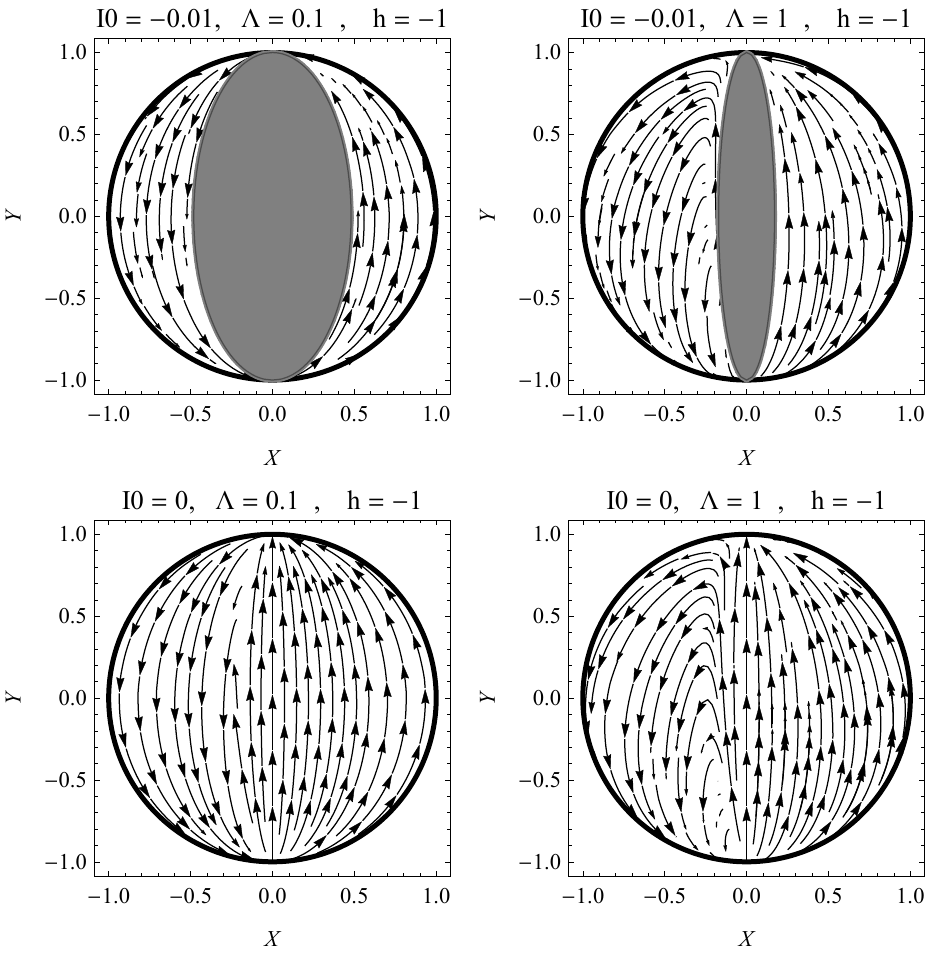}\caption{Phase-space
portraits for Szekeres system associated to \eqref{74} and \eqref{75} in the compactified variables $\left(
X,Y\right)$ for $I_{0}\leq 0$ and $\Lambda>0$ and $h=-1$ corresponding to \emph{Case a.2}.  The gray region is the forbidden  region $x^2 < -\frac{3 I_0}{\Lambda}$.}%
\label{fig9c}%
\end{figure}

In Fig.  \ref{fig9c} the phase-space
portraits for Szekeres system associated to \eqref{74} and \eqref{75} is showed in the compactified variables $\left(
X,Y\right)$ for $I_{0}\leq 0$ and $\Lambda>0$ and $h=-1$ corresponding to \emph{Case a.2}.  The gray region is the forbidden  region $x^2 \leq -\frac{3 I_0}{\Lambda}$.

\subsubsection{Case b:  $\Lambda<0, \; I_0>0, \; x^2 \leq -\frac{3 I_0}{\Lambda},\; y<0$.}

In this example the variable $x$ is bounded. Since $y<0$ we propose the bounded variable 
\begin{equation}
    Y= \frac{y}{y-1}  
\end{equation}
when $y\rightarrow  -\infty$, $Y\rightarrow 1$.

The dynamical system becomes 
\begin{align}
& \frac{dx}{d \bar{t}}= -3 I_0-\Lambda  x^2, \label{sz.B1}\\
& \frac{dY}{d \bar{t}}=\frac{3}{4} (Y-1)^2 (4 h x-1)+\Lambda  x
   Y (Y-1). \label{sz.B2} 
\end{align}
The stationary points at infinity are 
\[I_{-}: (x,Y)= \left(-\sqrt{3}
   \sqrt{-\frac{I_0}{\Lambda }} ,1\right)\]
 and   
\[I_{+}: (x,Y)= \left(\sqrt{3}
   \sqrt{-\frac{I_0}{\Lambda }} ,1\right)\]
   with eigenvalues
   $\left\{2 \sqrt{3} \Lambda  \sqrt{-\frac{I_0}{\Lambda }},-\sqrt{3}
   \Lambda  \sqrt{-\frac{I_0}{\Lambda }}\right\}$ and $\left\{-2 \sqrt{3} \Lambda  \sqrt{-\frac{I_0}{\Lambda }},\sqrt{3}
   \Lambda  \sqrt{-\frac{I_0}{\Lambda }}\right\}$, respectively. Hence, they are saddles because $\Lambda<0$ and $I_0>0$.

For $\Lambda<0$, the points at the finite region are  
\[Q_{-}: (x,y)= \left(\frac{\sqrt{3} \sqrt{-I_0 \Lambda }}{\Lambda },  \frac{144 h^2 I_0+48 h I_0 \Lambda
   +\Lambda  \left(4 \sqrt{3} \sqrt{-I_0 \Lambda }+3\right)}{16
   I_0 (3 h+\Lambda )^2+3 \Lambda }\right)\]
 and 
\[Q_{+}: (x,y)= \left(-\frac{\sqrt{3} \sqrt{-I_0 \Lambda }}{\Lambda }, \frac{144 h^2 I_0+48 h
   I_0 \Lambda +\Lambda  \left(3-4 \sqrt{3} \sqrt{-I_0 \Lambda
   }\right)}{16 I_0 (3 h+\Lambda )^2+3 \Lambda }\right).\]
   The eigenvalues of $Q_{\pm}$ are 
   $\left\{\pm 2 \sqrt{3} \sqrt{-I_0 \Lambda }, \pm \sqrt{3} \sqrt{-I_0
   \Lambda }\right\}$, respectively, which means that
point $Q_{-}$ is always a source and $Q_+$ is a sink. 

\begin{figure}[ptb]
\centering\includegraphics[width=0.9\textwidth]{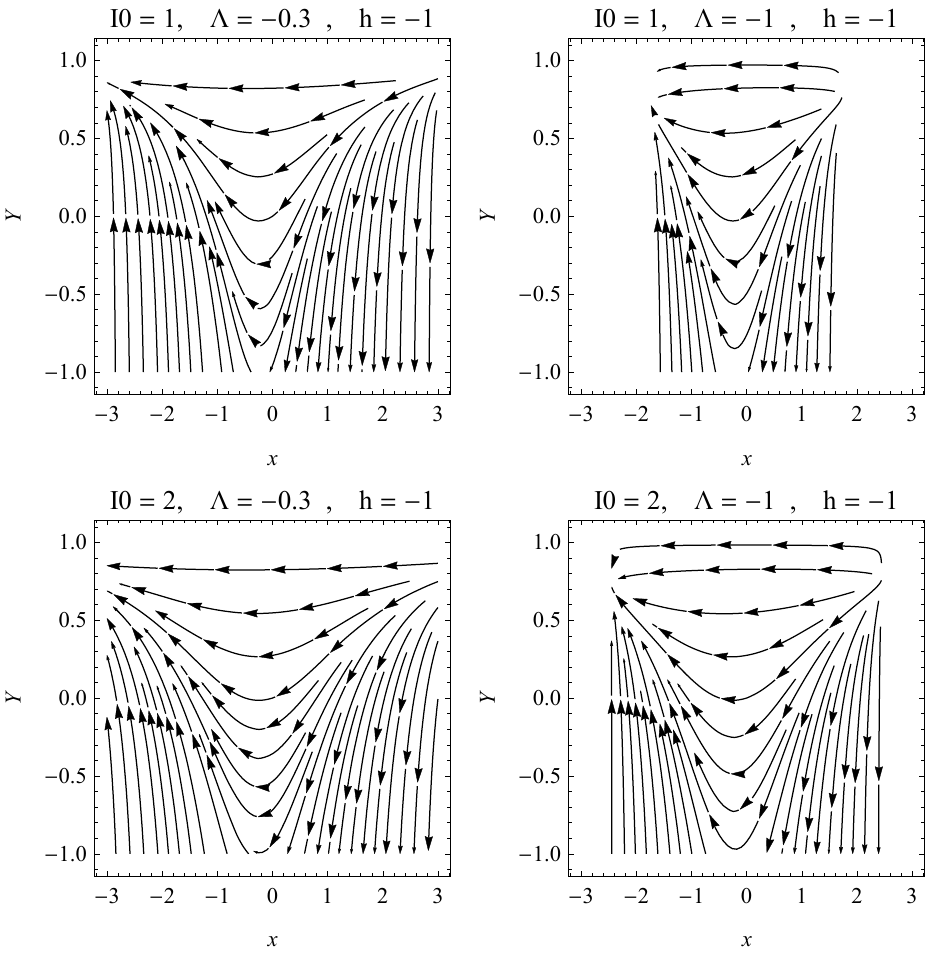}\caption{Phase-space portraits for the dynamical system \eqref{sz.B1}- \eqref{sz.B2}.}%
\label{figB}%
\end{figure}

In Fig. \ref{figB} some phase-space portraits for the dynamical system \eqref{sz.B1}- \eqref{sz.B2} are presented.

\section{Asymptotic behavior}

\label{sec5a}

In this section, we investigate the evolution of the kinematic and dynamical
variables for the field equations \eqref{ss1.01}- \eqref{ss1.05} by using
dimensionless variables. Such an approach has been widely studied in literature and can give us important information about the physical behavior
of the exact solutions at stationary points. The dynamics of the Szekeres
system in the $\theta$-normalization, known also as$~$Hubble normalization,
were studied by \cite{silent} for the case with zero-valued cosmological
constant. Recently, in \cite{as6} the case of a nonzero cosmological
constant was considered. However, in the $\theta$-normalization, the expansion
rate cannot change sign, that is, it cannot take the value $\theta=0$. But,
from the analysis in \cite{szek} and \cite{sz2} we know that because a family
of solutions describes Kantowski-Sachs universes, the expansion rate can 
change sign. 

Thus, in this work we define new dimensionless variables, different from that
of the $\theta$-normalization, say,
\begin{equation}
\omega_{D}=\frac{3\rho_{D}%
}{\left(  1+\theta^{2}\right)  },~ \Sigma=\frac{\sigma}{\sqrt{1+\theta^{2}}}, ~\alpha=\frac{E}{1+\theta^{2}},
\end{equation}%
\begin{equation}
\omega_{\Lambda}=\frac{\Lambda}{1+\theta^{2}},  ~\eta=\frac{\theta}{\sqrt{1+\theta^{2}}}, ~\omega_{R}=\frac{^{\left(  3\right)  }R}{\left(  1+\theta^{2}\right)  }.
\end{equation}

Therefore, the field equations read%
\begin{equation}
\omega_{D}^{\prime}=\frac{1}{3}\eta\omega_{D}\left(  2\eta^{2}-3+36\Sigma
^{2}+\omega_{D}-6\omega_{\Lambda}\right), \label{sz.10}%
\end{equation}%
\begin{equation}
\Sigma^{\prime}=\frac{1}{6}\left(  \Sigma\left(  2\eta^{3}+ 6\Sigma+\eta\left(
36\Sigma^{2}-4+\omega_{D}-6\omega_{\Lambda}\right)  \right)  - 6 \alpha\right)
,\label{sz.11}%
\end{equation}%
\begin{equation}
\alpha^{\prime}=\frac{1}{6}\left(  2\alpha\left(  2\eta^{3}-9\Sigma
+\eta\left(  36\Sigma^{2}-3+\omega_{D}-6\omega_{\Lambda}\right)  \right)
-\omega_{D}\Sigma\right), \label{sz.12}%
\end{equation}%
\begin{equation}
\omega_{\Lambda}^{\prime}=\frac{1}{3}\eta\omega_{\Lambda}\left(  2\eta
^{2}+36\Sigma^{2}+\omega_{D}-6\omega_{\Lambda}\right), \label{sz.13}%
\end{equation}%
\begin{equation}
\eta^{\prime}=\frac{1}{6}\left(  \eta^{2}-1\right)  \left(  2\eta^{2}%
+36\Sigma^{2}+\omega_{D}-6\omega_{\Lambda}\right), \label{sz.14}%
\end{equation}
with algebraic constraint%
\begin{equation}
\eta^{2}-9\Sigma^{2}-\omega_{D}+\frac{3}{2}\omega_{R}-3\omega_{\Lambda
}=0,\label{sz.15}%
\end{equation}
where we have defined the new derivative $f^{\prime}=\frac{1}{\sqrt{1+\theta^2}}\dot{f}$.

Every stationary point of the dynamical system \eqref{sz.10}-\eqref{sz.14} with constraint \eqref{sz.15}
describes an asymptotic solution for the inhomogeneous background space
\eqref{ss1.06}. The physical properties of the background spaces follow from
the values of the anisotropic parameter $\Sigma$ and of the curvature scalar
$\omega_{R}$. Indeed, for a stationary point with $\Sigma=0$ the background
space is isotropic, which means that the asymptotic solution at the point will
belong to the Friedmann-Lema\^{\i}tre-Robertson-Walker (-like) family of
solutions, while for $\Sigma\neq0$ to the Bianchi I, Kantowski-Sachs or to the
Bianchi III (-like) universes  depend on the sign of the curvature scalar. 
Moreover, $\eta>0$ remarks $\theta>0$ which correspond to an expanding
universe. While $\eta<0$, i.e. $\theta<0,$ corresponds to a shrinking universe.
Finally, $\eta=0$  describes static spacetimes with $\theta=0$. %

\begin{table}[tbp] \centering
\caption{The stationary points for the system \eqref{sz.10}-\eqref{sz.14} with constraint \eqref{sz.15}}%
\begin{tabular}
[c]{ccccc}\hline\hline
\textbf{Point} & $\left(  \omega_{D},\Sigma,\alpha,\omega_{\Lambda}%
,\eta,\omega_{R}\right)$ & \textbf{Spacetime} & \textbf{Eigenvalues} &
\textbf{Stability}\\\hline
$A_{1}^{\pm}$ & $\left(  1,0,0,0,\pm1,0\right)$ & Flat FLRW & $\pm1,\pm
1,\mp\frac{1}{2},\pm\frac{1}{3},\pm\frac{1}{3}$ & Saddle points\\
$A_{2}^{\pm}$ & $\left(  0,0,0,0,\pm1,-\frac{2}{3}\right)$ & Milne &
$\pm\frac{2}{3},\pm\frac{2}{3},\mp\frac{1}{3},\mp\frac{1}{3},\mp\frac{1}{3}$ &
Saddle points\\
$A_{3}^{\pm}$ & $\left(  0,\pm\frac{1}{3},\frac{2}{9},0,\pm1,0\right)$ &
Kasner & $\pm2,\pm2,\pm\frac{5}{3},\pm1,\pm\frac{2}{3}$ & $A_{3}^{+}$ source \\
&&&& $A_{3}^{-}$ attractor\\
$A_{4}^{\pm}$ & $\left(  0,\mp\frac{1}{12},\frac{1}{32},0,\pm1,-\frac{5}%
{8}\right)$ & Kantowski-Sachs & $\pm\frac{3}{4},\pm\frac{3}{4},\mp\frac
{5}{8},\mp\frac{1}{4},\pm\frac{1}{4}$ & Saddle points\\
$A_{5}^{\pm}$ & $\left(  0,\pm\frac{1}{3},0,0,\pm1,0\right)$ & Kasner &
$\pm2,\pm2,\pm2,\pm1,\pm1$ & $A_{5}^{+}$ source \\
&&&& $A_{5}^{-}$ attractor\\
$A_{6}^{\pm}$ & $\left(  0,\pm\frac{1}{6},0,0,\pm1,-\frac{1}{2}\right)$ &
Kantowski-Sachs & $\pm1,\pm1,\mp\frac{1}{2},\pm\frac{1}{2},0$ & Saddle
points\\
$A_{7}^{\pm}$ & $\left(  -3,\mp\frac{1}{3},\frac{1}{6},\pm1,0\right)$ & Not
physically  &  & \\
&& acceptable &  & \\
$B_{1}$ & $\left(  0,0,0,\frac{{\eta_0}^{2}}{3},{\eta_0},0\right)$ & de Sitter &
$-{\eta_0},-{\eta_0},-\frac{2}{3}{\eta_0},-\frac{2}{3}{\eta_0},0$ & CMT\\
$B_{2}$ & $\left(  0,-\frac{{\eta_0}}{3},\frac{{\eta_0}^{2}}{3},{\eta_0}^{2},{\eta_0}
,2{\eta_0}^{2}\right)$ & Bianchi III & $-2{\eta_0},-{\eta_0},-{\eta_0},{\eta_0},0$ & Saddle
point\\
$B_{3}$ & $\left(  0,\frac{2}{3}{\eta_0},0,3{\eta_0}^{2},{\eta_0},8{\eta_0}
^{2}\right)$ & Bianchi III & $-3{\eta_0},-2{\eta_0},-{\eta_0},2{\eta_0},0$ & Saddle point\\
$C_{1}$ & $\left(  \omega_{D},0,0,\frac{\omega_{D}}{6},0,\omega_{D}\right)$
& Closed FLRW & $-\sqrt{\frac{\omega_{D}}{6}},-\sqrt{\frac{\omega_{D}}{6}%
},\sqrt{\frac{\omega_{D}}{6}},\sqrt{\frac{\omega_{D}}{6}}\,,0$ & Saddle
point\\
$C_{2}$ & $\left(  0,0,0,0,0,0\right)$ & Minkowski & $0,0,0,0,0$ &
Unstable\\\hline\hline
\end{tabular}
\label{tab1}
\end{table}

Additionally, we study the stability properties of the stationary points. This
information is essential because we extract important information about the
evolution of the solution near the stationary points, as  we can also construct
the evolution of the cosmological history provided by the specific model.

The stability of stationary points for the system \eqref{sz.10}-\eqref{sz.14} with constraint \eqref{sz.15} for the following set of points $P=\left(  \omega_{D}, \Sigma, \alpha, \omega_{\Lambda},  \eta, \omega_{R}\right)$ is classified. We give the value of $\omega_{R}$, for stability analysis coordinates $\left(  \omega_{D}, \Sigma, \alpha, \omega_{\Lambda},  \eta\right)$ are considered. 
Points $A_{1}^{\pm}=\left(  1,0,0,0,\pm1,0\right)$ describe singular
solutions where only the pressureless matter fluid dominates. The background
space is isotropic with zero valued spatial curvature, which means that it is
reduced to the spatially flat Friedmann-Lema\^{\i}tre-Robertson-Walker
background space. Points $A_{2}^{\pm}=\left(  0,0,0,0,\pm1,-\frac{2}%
{3}\right)$  provide  asymptotic solutions in an isotropic background space
with negative curvature. Indeed, the exact solutions at the points are the
(inhomogeneous) Milne universes. The asymptotic solutions at the stationary
points $A_{3}^{\pm}=\left(  0,\pm\frac{1}{3},\frac{2}{9},0,\pm1,0\right)$ 
correspond to anisotropic universes with zero valued spatial curvature, that
is, they describe (inhomogeneous) Bianchi I vacuum solutions, i.e. Kasner
(-like) universes. Moreover, the background space at the stationary points
$A_{4}^{\pm}=\left(  0,\mp\frac{1}{12},\frac{1}{32},0,\pm1,-\frac{5}%
{8}\right)$ is anisotropic with negative valued spatial curvature, thus, the
solution is that of the vacuum Kantowski-Sachs (-like) spacetime. Furthermore,
inhomogeneous Kasner solutions are described also by the stationary points
$A_{5}^{\pm}=\left(  0,\pm\frac{1}{3},0,0,\pm1,0\right)  .$ The two of
stationary points $A_{6}^{\pm}=\left(  0,\pm\frac{1}{6},0,0,\pm1,-\frac{1}%
{2}\right)$ describe vacuum Kantowski-Sachs spacetimes. Moreover, the two
stationary points $A_{7}^{\pm}=\left(  -3,\mp\frac{1}{3},\frac{1}{6}%
,\pm1,0\right)$ are not physically accepted because they provide negative
energy density for the pressureless fluid source $\rho_{D}$.

These stationary points satisfy $\eta^{2}=1$ which
indicates that $\theta$ reaches infinite.\ Indeed the asymptotic solutions
described by the stationary points $A_{I}^{\pm},~I=1,2...7$, correspond to
singular solutions in which $\theta\propto t^{-1}$. \ There exist the
additional stationary points with ${\eta_0}^{2}\neq1~$which provide an exponential
expansion rate  $\theta\propto e^{H_{0}t}$. There are curves of stationary points,
$B_{1}=\left(  0,0,0,\frac{{\eta_0}^{2}}{3},{\eta_0},0\right)$, $B_{2}=\left(
0,-\frac{{\eta_0}}{3},\frac{{\eta_0}^{2}}{3},{\eta_0}^{2},{\eta_0},2{\eta_0}^{2}\right)$ and
$B_{3}=\left(  0,\frac{2}{3}{\eta_0},0,3{\eta_0}^{2},{\eta_0}, 8{\eta_0}^{2}\right)
$. The curve of stationary points $B_{1}$  describes the asymptotic solution for a spatially
flat Friedmann-Lema\^{\i}tre-Robertson-Walker (FLRW) (-like) spacetime
dominated by the cosmological constant, that is, it describes de Sitter
universe. One the other hand, the curves of stationary points $B_{2}$ and $B_{3}$ describe Bianchi
III (-like) spacetimes. 

Finally, there exist two lines of stationary points with $\eta=0$, they are
$C_{1}=\left(  \omega_{D},0,0,\frac{\omega_{D}}{6},0,\omega_{D}\right)$ and
$C_{2}=\left(  -18\Sigma^{2}, \Sigma,\Sigma^{2},3\Sigma^{2},0,0\right)$. The
asymptotic solution at the line of stationary points $C_{1}$ describes an isotropic
universe with positive spatially curvature, that is, it describes the closed
Friedmann-Lema\^{\i}tre-Robertson-Walker (-like) spacetime and exists only
for positive cosmological constant. Furthermore, points over line $C_{2}$ are physically
accepted only when $\Sigma=0$, and in such case we have the physical stationary point $C_{2}=\left(
0,0,0,0,0,0\right)$. Therefore, $C_{2}$ describes the Minkowski solution. It
is very interesting that there are two stationary points in which the
expansion rate $\theta$ changes sign, and the trajectories for the field
equations can go from the region which describes an expanding universe,
$\theta>0$, to the region which the universe shrinks, i.e. $\theta<0$. 
In Table \ref{tab1},  the above discussion is summarized. Additionally, 
the eigenvalues around the stationary points for the linearized system \eqref{sz.10}- \eqref{sz.14} 
are presented. From the eigenvalues, we infer that points $A_{1}^{\pm}%
$, $A_{2}^{\pm}$, $A_{4}^{\pm}$,~$B_{2}$,~$B_{3}\,\ $ and $C_{1}$ are always
saddle points. Meanwhile, Points $A_{3}^{+}$, $A_{5}^{+}$ are sources describing
unstable solutions, while $A_{3}^{-}$ and $A_{5}^{-}$ are attractors. The line of stationary points
$B_{1}$ has five negative eigenvalues and a zero eigenvalue. Thus, it has a
4D stable manifold and a  1D center manifold, which can be determined by applying the center manifold
theorem (CMT). As far as to the line of stationary points $C_{2}$ is concerned the five eigenvalues are
all zero. However, from the phase-space diagrams presented below it follows
that the point $C_{2}$ always describes the unstable Minkowski universe.

\begin{figure}[ptb]
\centering\includegraphics[width=0.9\textwidth]{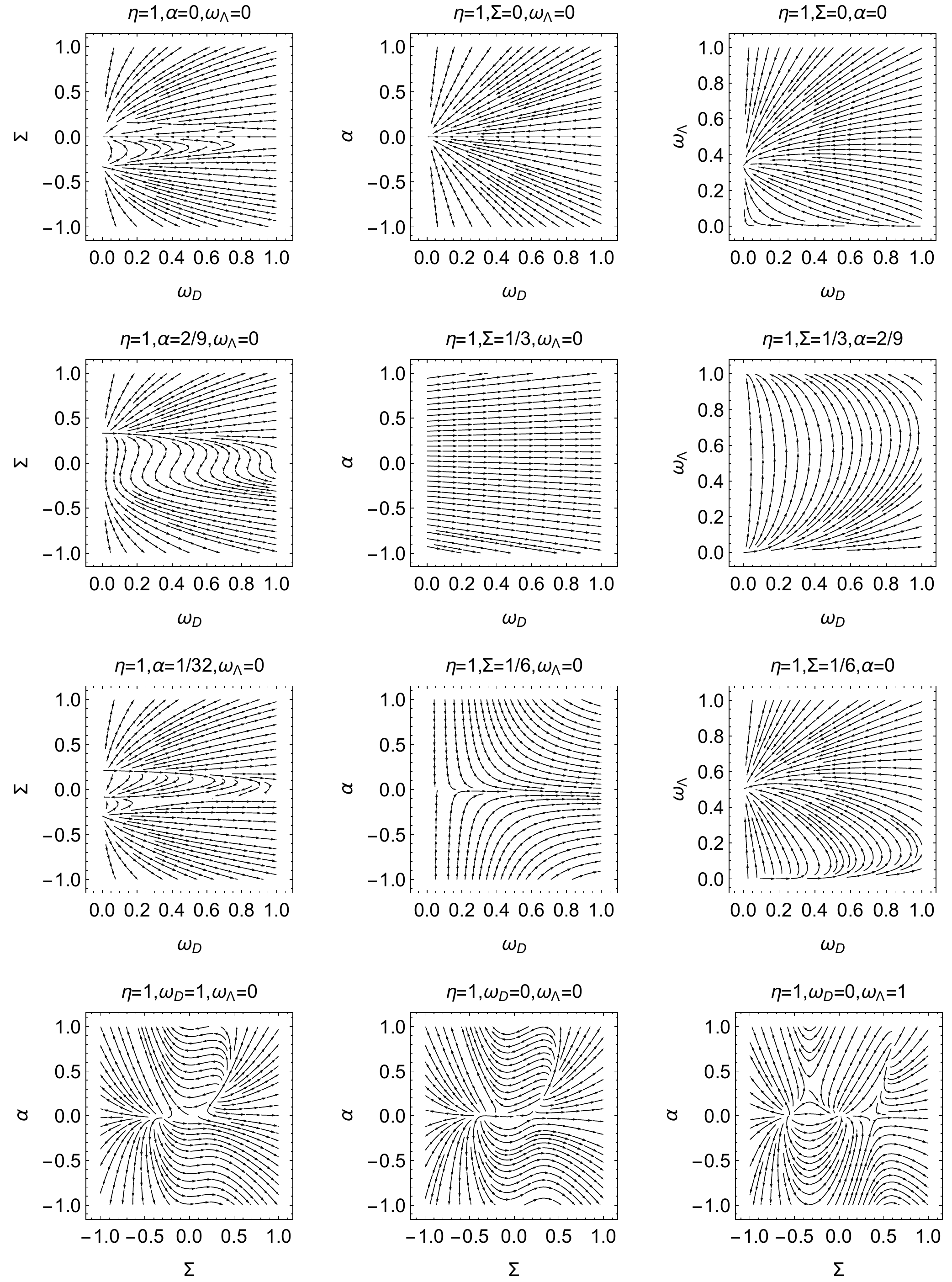}\caption{Phase-space
portraits for the dynamical system \eqref{sz.10}- \eqref{sz.14} in the
two-dimensional planes $\left(  \omega_{D},\Sigma\right)  \,$, $\left(
\omega_{D},\alpha\right)$, $\left(  \omega_{D},\omega_{\Lambda}\right)$,
$\left(  \Sigma,\alpha\right)$ for $\eta=1$.}%
\label{fig11}%
\end{figure}

\begin{figure}[ptb]
\centering\includegraphics[width=0.9\textwidth]{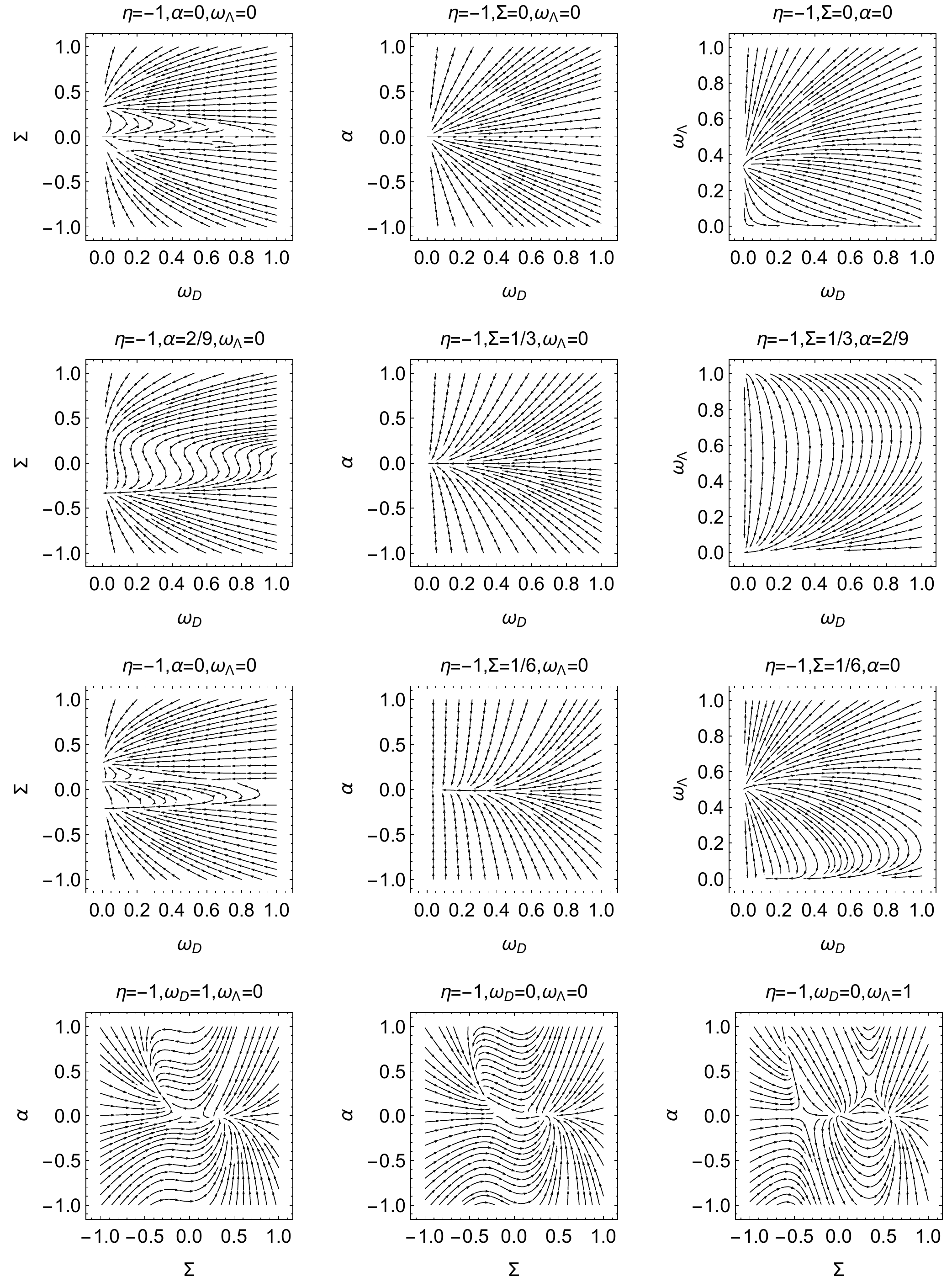}\caption{Phase-space
portraits for the dynamical system \eqref{sz.10}- \eqref{sz.14} in the
two-dimensional planes $\left(  \omega_{D},\Sigma\right)  \,$, $\left(
\omega_{D},\alpha\right)$, $\left(  \omega_{D},\omega_{\Lambda}\right)$,
$\left(  \Sigma,\alpha\right)$ for $\eta=-1$.}%
\label{fig12}%
\end{figure}

\begin{figure}[ptb]
\centering\includegraphics[width=1\textwidth]{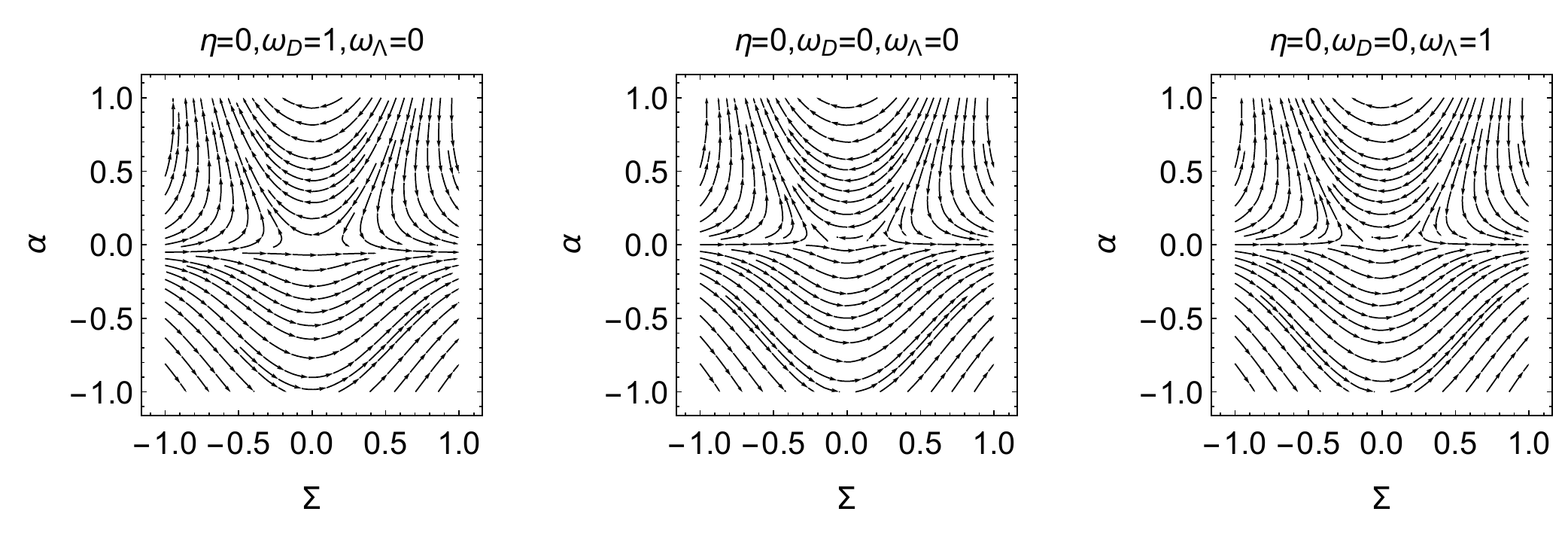}\caption{Phase-space
portraits for the dynamical system \eqref{sz.10}- \eqref{sz.14} in the
two-dimensional plane $\left(  \Sigma,\alpha\right)$ for $\eta=0$.}%
\label{fig13}%
\end{figure}

\begin{figure}[ptb]
\centering\includegraphics[width=0.9\textwidth]{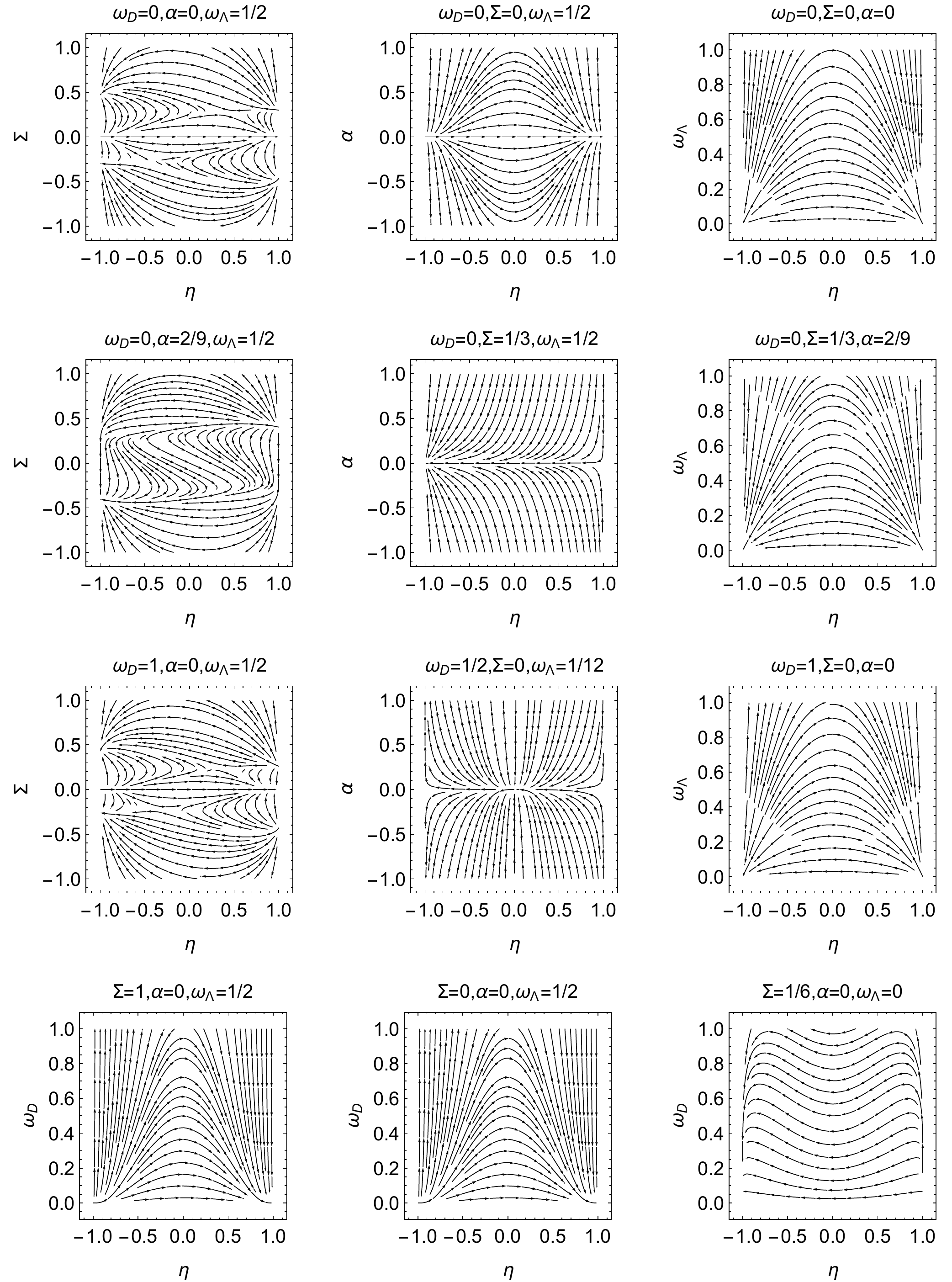}\caption{Phase-space
portraits for the dynamical system \eqref{sz.10}- \eqref{sz.14} in the
two-dimensional planes $\left(  \eta,\Sigma\right), \left(  \eta
,\alpha\right)$, $\left(  \eta,\omega_{\Lambda}\right)$ and $\left(
\eta,\omega_{D}\right)  .$}%
\label{fig14}%
\end{figure}

In Figs. \ref{fig11}, \ref{fig12}, \ref{fig13} and \ref{fig14} we present
phase-space portraits for the field equations \eqref{sz.10}- \eqref{sz.14} in
the two-dimensional planes $\left(  \omega_{D},\Sigma\right)  \,$, $\left(
\omega_{D},\alpha\right)$, $~\left(  \omega_{D},\omega_{\Lambda}\right)
$,~$\left(  \Sigma,\alpha\right)$ for $\eta=1$ (Fig. \ref{fig11}), $\eta=-1$
(Fig. \ref{fig12}) and $\eta=0$ (Fig. \ref{fig13}). In Fig. \ref{fig14} the
phase-space portraits are in the two-dimensional planes $\left(  \eta
,\Sigma\right), \left(  \eta,\alpha\right)$, $\left(  \eta,\omega_{\Lambda
}\right)$ and $\left(  \eta,\omega_{D}\right).$ 

The set of stationary points $B_1$ can be parametrized as the curve $s(\eta)= \left(  0,0,0,\frac{{\eta}^{2}}{3},{\eta},0\right)$.
Fixing a point of $B_{1}$ with $\eta=\eta_0$ the eigensystem at the stationary point fixed is given by \newline 
$\left\{
\begin{array}{ccccc}
 0 & -\eta_0  & -\eta_0  & -\frac{2 \eta_0 }{3} & -\frac{2 \eta_0 }{3} \\
 \left(0,0,0,\frac{2 \eta_0 }{3},1\right) & \left(-\frac{2 \eta_0 }{\eta_0
   ^2-1},0,0,\frac{2 \eta_0 ^3}{3 \left(\eta_0 ^2-1\right)},1\right) &
   \left(0,\frac{3}{\eta_0 },1,0,0\right) & \left(0,0,0,\frac{2 \eta_0
   ^3}{3 \left(\eta_0 ^2-1\right)},1\right) & (0,1,0,0,0) \\
\end{array}
\right\}.$
Since the eivenvector associated to the zero eigenvalue, $ \left(0,0,0,\frac{2 \eta_0 }{3},1\right)$ is parallel to the tangent vector
$v= \frac{d s(\eta)}{d \eta}|_{\eta=\eta_0}= \left(0,0,0,\frac{2 \eta_0 }{3},1\right)$ at the given point with coordinates $\left(0,0,0,\frac{{\eta_0}^{2}}{3},{\eta_0},0\right)$ in the curve $B_1$. Then, the line of stationary points  $B_{1}$ is normally hyperbolic. A set of non-isolated stationary points is said to be normally hyperbolic if the only eigenvalues with zero real parts are those whose corresponding eigenvectors are tangent to the set.
Since by definition any point on a set of non-isolated stationary points will have at least
one eigenvalue which is zero, all points in the set are non-hyperbolic. However, a set that is
normally hyperbolic can be completely classified as per its stability by considering the signs of eigenvalues in the remaining directions (i.e., for a curve, in the
remaining $n-1$ directions) (see \cite{aulbach}, pp. 36).
\newline
When these arguments are applied to $B_1$, it follows that the line of stationary points is stable. 

Another way to see this is using the CMT. 

Defining the similarity matrix
\begin{equation}
    S= \left(
\begin{array}{ccccc}
 0 & -\frac{2 \eta_0}{\eta_0^2-1} & 0 & 0 & 0 \\
 0 & 0 & \frac{3}{\eta_0} & 0 & 1 \\
 0 & 0 & 1 & 0 & 0 \\
 \frac{2 \eta_0}{3} & \frac{2 \eta_0^3}{3
   \left(\eta_0^2-1\right)} & 0 & \frac{2  \eta_{0}^3}{3 \left(\eta_0^2-1\right)} & 0 \\
 1 & 1 & 0 & 1 & 0 \\
\end{array}
\right),
\end{equation}
and the linear transformation 
\begin{equation}
\mathbf{X}:=\left(x_1, x_2, x_3,x_4, x_5\right)= S^{-1} \cdot \left( \omega_{D},\Sigma ,\alpha ,\omega_\Lambda
   -\frac{\eta_0^2}{3},\eta -\eta_0\right)
\end{equation}
such that 
\begin{small}
\begin{align}
  & \left( \omega _{D}, \Sigma, \alpha, \omega_\Lambda, \eta \right)= \nonumber \\
  & \left(-\frac{2 \eta_0
   x_2}{\eta_0^2-1},  \frac{3
   x_3}{\eta_0}+x_5,
   x_3,  \frac{\eta_0
   \left(\eta_0^3-\eta_0+2 \eta_0^2
   (x_1+x_2+x_4)-2 x_1\right)}{3
   \left(\eta_0^2-1\right)},  \eta_0+x_1+x_2+x_4\right).
\end{align}
\end{small}
This transformation translates the point with coordinates $\left(0,0,0,\frac{{\eta_0}^{2}}{3},{\eta_0},0\right)$ at the curve $B_1$  to the origin. 
The new system can be symbolically written as 
\begin{align}
 \left(\begin{array}{c}
x_1^{\prime} \\
x_2^{\prime} \\
x_3^{\prime} \\
x_4^{\prime} \\
x_5^{\prime} \\
\end{array}
\right)=  \underbrace{\left(
\begin{array}{ccccc}
 0 & 0 & 0 & 0 & 0 \\
 0 & -\eta_0  & 0 & 0 & 0 \\
 0 & 0 & -\eta_0  & 0 & 0 \\
 0 & 0 & 0 & -\frac{2 \eta_0 }{3} & 0 \\
 0 & 0 & 0 & 0 & -\frac{2 \eta_0 }{3} \\
\end{array}
\right) }_{=\mathbf{J}}\left(\begin{array}{c}
x_1 \\
x_2   \\
x_3 \\
x_4  \\
x_5  \\
\end{array}
\right) + \left(\begin{array}{c}
g_1(\mathbf{X}) \\
g_2(\mathbf{X}) \\
g_3(\mathbf{X}) \\
g_4(\mathbf{X}) \\
g_5(\mathbf{X}) \\
\end{array}
\right), \label{eq90}
\end{align}
or 
\begin{equation}
    \mathbf{X}^{\prime}= \mathbf{J}  \mathbf{X} + \mathbf{g}(\mathbf{X})
\end{equation}
where the vector $\mathbf{g}(\mathbf{X})$ contains the nonlinear terms.
Hence, the local center manifold of the origin  is given locally by the graph 
\begin{align}
   \Big\{\mathbf{X}\in \mathbb{R}^5 &   : x_2=h_2(x_1), x_3=h_3(x_1),  x_4=h_4(x_1), x_5=h_5(x_1), \nonumber\\
    & h_2'(0)= h_3'(0)=h_4'(0)=h_5'(0)=0, \nonumber\\
    & h_2(0)=h_3(0)=h_4(0)=h_5(0)=0, |x_1|<\delta \Big\}
\end{align}
for $\delta$ small enough.
From the invariance of the center manifold of the origin for the flow of the dynamical systems it follows that $h_i$ satisfy the set of differential equations
\begingroup\makeatletter\def\f@size{10}\check@mathfonts
\begin{subequations}
\label{quasi_linearODE}
\begin{align}
& \eta_0 h_2 (x_1) - g_2(x_1, h_2(x_1),h_3(x_1),h_4(x_1),h_5(x_1)) + h_2'(x_1) g_1(x_1, h_2(x_1),h_3(x_1),h_4(x_1),h_5(x_1))=0, \label{93a}\\
& \eta_0 h_3 (x_1) - g_3(x_1, h_2(x_1),h_3(x_1),h_4(x_1),h_5(x_1)) + h_3'(x_1) g_1(x_1, h_2(x_1),h_3(x_1),h_4(x_1),h_5(x_1))=0, \label{93b}\\
& \frac{2}{3}\eta_0 h_4 (x_1) - g_4(x_1, h_2(x_1),h_3(x_1),h_4(x_1),h_5(x_1)) + h_4'(x_1) g_1(x_1, h_2(x_1),h_3(x_1),h_4(x_1),h_5(x_1))=0, \label{93c}\\
& \frac{2}{3}\eta_0 h_5 (x_1) - g_5(x_1, h_2(x_1),h_3(x_1),h_4(x_1),h_5(x_1)) + h_5'(x_1) g_1(x_1, h_2(x_1),h_3(x_1),h_4(x_1),h_5(x_1))=0. \label{93d}
\end{align}
\end{subequations}
\endgroup
Substituting  the ansatz 
\begin{equation}
    h_i(x_1) = \sum_{k=2}^{N-1} a_{i k} x_1^k + \mathcal{O}(x_1^N), i=2,3,4,5, 
\end{equation}
in eqs. \eqref{quasi_linearODE} and comparing coefficients of the same powers in $x_1$ we obtain  
$a_{22}= 0,a_{23}= 0,a_{24}= 0,a_{25}= 0,a_{26}= 0,a_{27}= 0,a_{28}= 0,a_{29}= 0,a_{32}= 0,a_{33}= 0,a_{34}= 0,a_{35}= 0,a_{36}= 0,a_{37}= 0,a_{38}= 0,a_{39}= 0,a_{42}=
   \frac{\eta_0^2-1}{2 \eta_0},a_{43}= \frac{\left(\eta_0^2-1\right)^2}{2 \eta_0^2},a_{44}= \frac{5 \left(\eta_0^2-1\right)^3}{8 \eta_0^3},a_{45}= \frac{7 \left(\eta_0^2-1\right)^4}{8  \eta_{0}^4},a_{46}= \frac{21 \left(\eta_0^2-1\right)^5}{16 \eta_0^5},a_{47}= \frac{33 \left(\eta_0^2-1\right)^6}{16 \eta_0^6},a_{48}= \frac{429 \left(\eta_0^2-1\right)^7}{128 \eta_0^7},a_{49}= \frac{715
   \left(\eta_0^2-1\right)^8}{128 \eta_0^8},a_{52}= 0,a_{53}= 0,a_{54}= 0,a_{55}= 0,a_{56}= 0,a_{57}= 0,a_{58}= 0,a_{59}= 0$ for $N=10$. 
This is our induction start.  Assuming  $a_{2 k}= a_{3 k}= a_{5 k}=0, \forall k,  2 \leq k\leq N-1, N\geq 10$ and equating to zero the coefficients of all orders of $x_i^k$ up to order $N+1$, we obtain $a_{2 N}= a_{3 N}= a_{5 N}=0$. That is, assuming $h_i(x_1) = a_{i N} x_1^N + \mathcal{O}(x_1^{N+1}), i=2,3,5$, 
we obtain from \eqref{93a}, \eqref{93b} and \eqref{93d} (and by neglecting the terms $\mathcal{O}(x_1^{N+1})$) that 
\begin{align}
& a_{2 N} P(x_1, h_4(x_1))   = 0, a_{3 N} P(x_1, h_4(x_1))   = 0, \label{eq96}\\
&a_{3N} Q(x_1, h_4(x_1))+a_{5N} R(x_1, h_4(x_1))=0, \label{eq97}
\end{align}
where
\newline
$P(x_1, h_4)= \eta_0^2 \left(\eta_0^2-1\right) N h_4^4+\eta_0 h_4^2 \left(\left(\eta_0^2-3\right) x_1 \left(3 \left(\eta_0^2-1\right) N+2\right)-2 \eta_0 \left(\eta_0^2-1\right)
   N\right)+h_4^3 \left(\eta_0 \left(\eta_0^4-4 \eta_0^2+1\right) N+2 \left(\eta_0^2-1\right) x_1 \left(\left(2 \eta_0^2-1\right) N+1\right)\right)+x_1 h_4 \left(-2 \eta_0^4 N+ \eta_{0}^2 (6 N-7)+3\right)-3 \eta_0 \left(\eta_0^2-1\right) x_1$,  \newline 
   $Q(x_1, h_4)= \left(-3 \left(\eta_0^2-1\right) x_1 h_4^3-3 \eta_0 \left(\eta_0^2-3\right) x_1 h_4^2+\left(9 \eta_0^2-3\right) x_1 h_4\right)$, \newline
   $R(x_1, h_4)= \eta_0^2
   h_4^2 \left(\left(\eta_0^2-3\right) x_1 \left(3 \left(\eta_0^2-1\right) N+1\right)-2 \eta_0 \left(\eta_0^2-1\right) N\right)+\eta_0 h_4^3 \left(\eta_0 \left(\eta_0^4-4 \eta_{0}^2+1\right) N+\left(\eta_0^2-1\right) x_1 \left(\left(4 \eta_0^2-2\right) N+1\right)\right)+\eta_0^3 \left(\eta_0^2-1\right) N h_4^4+h_4 \left(2 \eta_0 x_1-2 \eta_0^3 x_1
   \left(\left(\eta_0^2-3\right) N+2\right)\right)-2 \eta_0^2 \left(\eta_0^2-1\right) x_1$. \newline
   On the other hand, the function 
   \begin{equation}
       h_4(x_1)=-x_1+\frac{\eta_0-\sqrt{\eta_0 \left(\eta_0-2 \eta_0^2 x_1+2 x_1\right)}}{\eta_0^2-1}\label{eq98}
   \end{equation}
   satisfies $h_4(0)=h_4'(0)$ and  its series expansions around $x_1=0$ has exactly the first eight coefficients $a_{42}, \ldots a_{4 9}$ deduced at the inducting start. Substituting \eqref{eq98} in \eqref{eq96} and \eqref{eq97} we obtain $P(x_1, h_4(x_1))\neq 0, Q(x_1, h_4(x_1))\neq 0$ and $R(x_1, h_4(x_1))\neq 0$ for $x_1\neq 0$ and $0<\eta_0<1$, which implies $a_{2 N}= a_{3 N}= a_{5 N}=0$. Using induction over $N$ we obtain the exact solution
\begin{align}
  h_2(x_1)\equiv 0,  h_3(x_1)\equiv 0,    h_4(x_1)=-x_1+\frac{\eta_0-\sqrt{\eta_0 \left(\eta_0-2 \eta_0^2 x_1+2 x_1\right)}}{\eta_0^2-1},  h_5(x_1)\equiv 0.  
\end{align}
That is, the local center manifold of the origin is 
\begin{align}
   \Big\{\mathbf{X}\in \mathbb{R}^5 &   : x_2=0, x_3=0,  x_4=-x_1+\frac{\eta_0-\sqrt{\eta_0 \left(\eta_0-2 \eta_0^2 x_1+2 x_1\right)}}{\eta_0^2-1}, x_5=0, |x_1|<\delta \Big\}
\end{align}
and $x_1^{\prime}=0$ at the center manifold. 
In the original variables the center manifold is 
\begin{align}
    \Big\{ \left(  \omega_{D},\Sigma,\alpha,\omega_{\Lambda},\eta\right) \in \mathbb{R}^5 & :     \omega_{D}= 0,\Sigma = 0,\alpha= 0, \nonumber \\
    & \omega_\Lambda= \frac{\eta_0 \left(\eta_0^5+\eta_0-2 \left(\eta_0^2-1\right) x_1-2 \eta_0^{5/2} \sqrt{\eta_0-2 \eta_0^2 x_1+2 x_1}\right)}{3
   \left(\eta_0^2-1\right)^2}, \nonumber \\
   & \eta = \frac{\eta_0^3-\sqrt{\eta_0 \left(\eta_0-2 \left(\eta_0^2-1\right) x_1\right)}}{\eta_0^2-1}, |x_1|\leq \delta\Big\}.
\end{align}
\begin{figure}[t!]
\centering\includegraphics[width=0.5\textwidth]{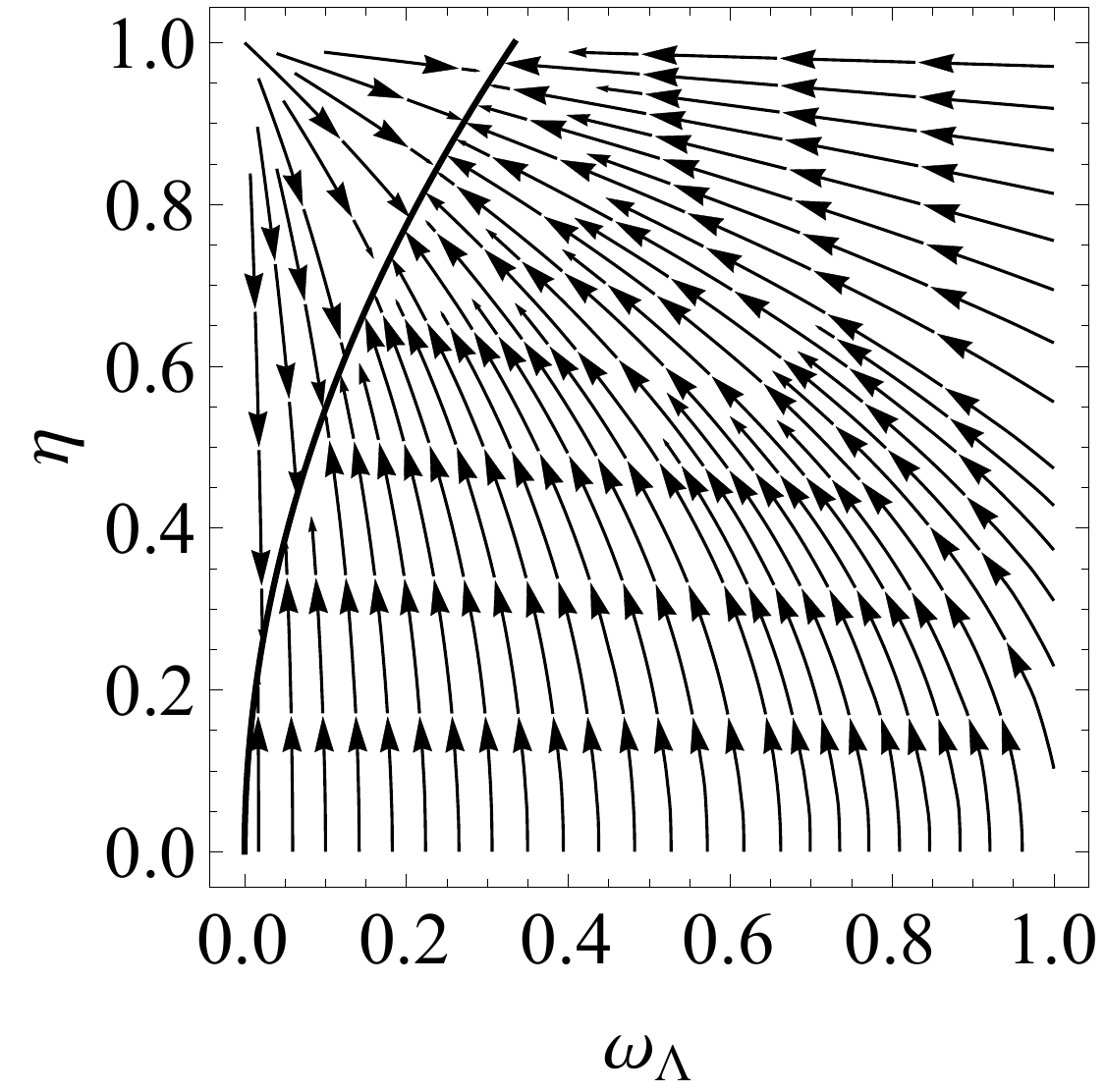}\caption{Phase portrait of the dynamical system \eqref{100A} - \eqref{100B} in the two-dimensional plane $(\omega_\Lambda, \eta)$ where  the stability of the de Sitter line of points $B_1:  \eta ^2-3 \omega_\Lambda=0$ (represented by a solid line) it is shown.}%
\label{figC1}%
\end{figure}
In our approximation we  have
\begin{align}
  \omega_\Lambda& =  \frac{\eta_0^2}{3}+\frac{2 \eta_0 x_1}{3}+\frac{\eta_0^2 x_1^2}{3}+\frac{1}{3} \eta_0 \left(\eta_0^2-1\right) x_1^3+\frac{5}{12} \left(\eta_0^2-1\right)^2 x_1^4+\frac{7 \left(\eta_0^2-1\right)^3
   x_1^5}{12 \eta_0} \nonumber\\
   & +\frac{7 \left(\eta_0^2-1\right)^4 x_1^6}{8 \eta_0^2}+\frac{11 \left(\eta_0^2-1\right)^5 x_1^7}{8 \eta_0^3}+\frac{143 \left(\eta_0^2-1\right)^6 x_1^8}{64 \eta_0^4}+\frac{715
   \left(\eta_0^2-1\right)^7 x_1^9}{192 \eta_0^5}+O\left(x_1^{10}\right),\\
   & \eta= \eta_0+x_1+\frac{\left(\eta_0^2-1\right) x_1^2}{2 \eta_0}+\frac{\left(\eta_0^2-1\right)^2 x_1^3}{2 \eta_0^2}+\frac{5
   \left(\eta_0^2-1\right)^3 x_1^4}{8 \eta_0^3}+\frac{7 \left(\eta_0^2-1\right)^4 x_1^5}{8 \eta_0^4} \nonumber\\
   & +\frac{21 \left(\eta_0^2-1\right)^5 x_1^6}{16 \eta_0^5}+\frac{33 \left(\eta_0^2-1\right)^6 x_1^7}{16
   \eta_0^6}+\frac{429 \left(\eta_0^2-1\right)^7 x_1^8}{128 \eta_0^7}+\frac{715 \left(\eta_0^2-1\right)^8 x_1^9}{128 \eta_0^8}+O\left(x_1^{10}\right).
\end{align}
Using these results we obtain that the center manifold is contained in the plane $(\omega_{\Lambda}, \eta)$ whose dynamics is given by 
\begin{align}
   & \omega_\Lambda^{\prime}= \frac{2}{3} \eta  \omega_\Lambda  \left(\eta ^2-3 \omega_\Lambda \right), \label{100A}\\
   &\eta^{\prime}= \frac{1}{3} \left(\eta ^2-1\right) \left(\eta ^2-3 \omega_\Lambda \right),\label{100B}
\end{align}
with constraint 
\begin{equation}
-\eta ^2+3 \omega_\Lambda -\frac{3  \omega_{R}}{2}=0.
\end{equation}
Then, the dynamics of the center manifold of $B_1$ can be inferred from the dynamics of the system \eqref{100A} - \eqref{100B}. 
The stability of the center manifold is illustrated in Fig.  \ref{figC1} where a phase portrait of the dynamical system \eqref{100A} - \eqref{100B} in the two-dimensional plane $(\omega_\Lambda, \eta)$ is displayed. Then, the stability of the de Sitter line of points $B_1:  \eta ^2-3 \omega_\Lambda=0$ it is shown.
That is, we show that the attractor at the finite regime in $\mathbb{R}^5$ is related with the de Sitter universe for a positive cosmological constant and positive $\eta$ (positive $H$).

\section{Conclusions}

\label{sec5}

In this study, we investigated the global dynamics for the Szekeres system
with a nonzero cosmological constant term. The Szekeres system has consisted
of an algebraic equation and four first-order ordinary differential
equations defined in the $\mathbb{R}^{4}$. The system of ordinary
differential equations is autonomous and admits a point-like Lagrangian.
Consequently, a Hamiltonian function exists and a sufficient number of
conservation laws from where infer the integrability properties of the
Szekeres system. By applying the Hamilton-Jacobi theory, we can reduce
the Szekeres system into a system of two first-order ordinary differential
equations.

Hence in the $\mathbb{R}^{2}$ space, we investigate the global dynamics for
the evolution of the Szekeres system in the finite and infinity regime. We
find that various values of the conservation laws, that is, of the free
variables the evolution of the Szekeres system differs. Moreover, we
investigate the dynamics on specific surfaces of special interests.

Finally, for completeness of our analysis, we study the evolution of the
physical variables of the Szekeres system with the nonzero cosmological
constant term, by using dimensionless variables different from that of the
Hubble-normalization. In the new variables the stationary points of the
Szekeres system determined while the stability properties are provided.

In the presence of the cosmological constant, the evolution of the Szekeres
system is different from that which was studied before when $\Lambda =0$. We
see that for $\Lambda >0$, there can exist an attractor in the finite regime
which describes the de Sitter universe.

\begin{acknowledgments}
This research was funded by  Agencia Nacional de Investigaci\'on y Desarrollo- ANID  through the program FONDECYT Iniciaci\'on grant no.
11180126 and by Vicerrector\'{\i}a de Investigaci\'on y Desarrollo Tecnol\'ogico at Universidad Cat\'olica del Norte. 
 Ellen de Los Milagros Fern\'andez Flores is acknowledged for proofreading this manuscript and for improving the English. 
\end{acknowledgments}

\appendix

\section{Szekeres system with $\Lambda=0$}

\label{app1}

In this Appendix we discuss about the global dynamics for Szekeres
system with $\Lambda=0$. Such analysis has been published before for other set
of variables in \cite{as7}. However, for the completeness of our analysis we
summarize the main results in the following lines. The dynamical system
\eqref{sz.01}, \eqref{sz.02} for $\Lambda=0$ reads%
\begin{align}
\dot{x}  &  =-\sqrt{3}y\left(  3I_{0}y+6\right), \label{A.1}\\
\dot{y}  &  =-\sqrt{3}\left(  x\left(  y^{3}-3\right)  +3hy^{2}\right). \label{A.2}
\end{align}
\begin{figure}[ptb]
\centering\includegraphics[width=1\textwidth]{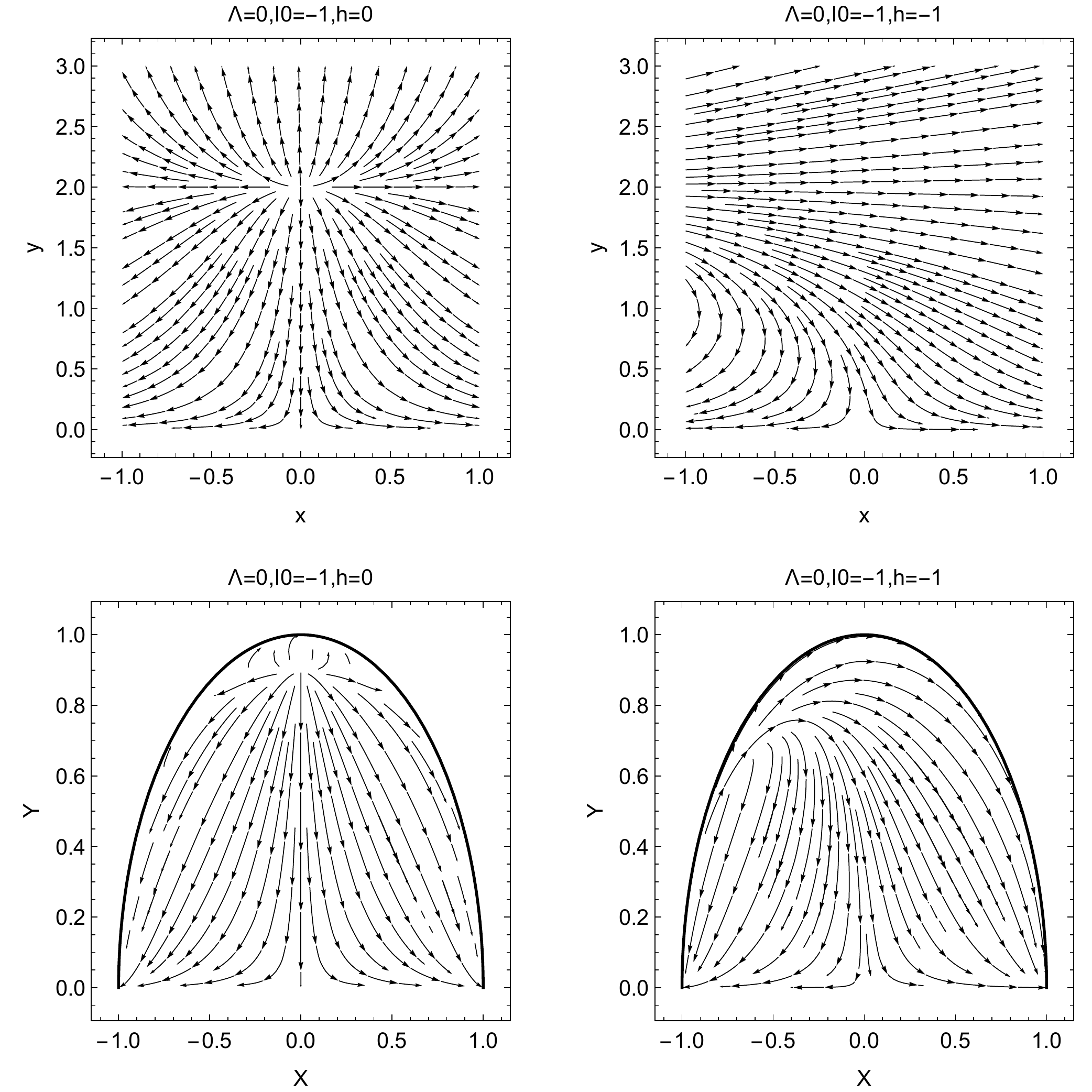}\caption{Phase-space
portraits for Szekeres associated to system \eqref{A.1}, \eqref{A.2} for different values of the free parameters
$I_{0}$ and $h$ and zero value for the cosmological constant, i.e. $\Lambda=0$.
Figures of the first row are in the local variables $\left(  x,y\right)$,
while figures of the second row are for the compactified variabels $\left(
X,Y\right)$. }%
\label{fig10}%
\end{figure}
The latter dynamical system in the finite regime admits the stationary points
$P_{0}^{\left( \Lambda=0\right)}=\left(  0,0\right)$ and $P_{1}^{\left(
\Lambda=0\right)  }=\left(  \frac{4h}{I_{0}^{2}},-\frac{2}{I_{0}}\right)$
which exist for $I_{0}\neq0$ and it is physically accepted when $I_{0}<0,$
because $y\geq0.$ The eigenvalues of the linearized system around point
$P_{0}^{\left( \Lambda=0\right)}$  are $-6\sqrt{3}$ and $3\sqrt{3}$, which
means that the point is saddle. Similarly, for point $P_{1}^{\left(
\Lambda=0\right)  }$ we derive the eigenvalues $6\sqrt{3}$ and $3\sqrt{3}$
which means that $P_{1}^{\left( \Lambda=0\right)}$ is a source. Thus, do not exists stationary point at the finite regime.

As far as the analysis at infinity is concerned, we can easily conclude from the
results of Section \ref{sec4} that results for
$\Lambda<0$ and $I_{0}<0$ apply also when  $\Lambda=0$ and $I_{0}<0$. Therefore, the trajectories can end
at infinity, while they are originated at infinity or the source point
$P_{1}^{\left(\Lambda=0\right)}$.

In Fig. \ref{fig10}  phase-space portraits for the Szekeres system with zero cosmological
constant terms for different values of the free variables are presented. We observe that the trajectories of the dynamical systems end at infinity.


\begin{thebibliography}{99}                                                                                               %


\bibitem {kras}A. Krasi\'{n}ski, \textit{Inhomogeneous Cosmological Models,
}Cambridge U.P., Cambridge, (1997)

\bibitem {szek}P. Szekeres, A class of inhomogeneous cosmological models,
Commun. Math. Phys. 41, 55 (1975) 10.1007/BF01608547

\bibitem {sz2}J.D. Barrow and J. Stein-Schabes, Inhomogeneous cosmologies with
cosmological constant, Phys. Lett. A 103, 315 (1984) 10.1016/0375-9601(84)90467-5

\bibitem {sz2a}G.M. Covarrubias, A class of Szekeres space-times with
cosmological constant, Astroph.\ Sp. Sci. 103, 401 (1984) 10.1007/BF00653757

\bibitem {lesame}R. Maartens, W. Lesame and G.F.R. Ellis, Consistency of dust
solutions with div H=0, Phys.Rev. D \textbf{55,} 5219 (1997) 10.1103/PhysRevD.55.5219

\bibitem {silent}M. Bruni, S.\ Matarrese and P. Ornella, Dynamics of Silent
Universes, Astroph. J. 445, 958 (1995) 10.1086/175755

\bibitem {sz1}D.A. Szafron, Inhomogeneous cosmologies: New exact solutions and
their evolution, J. Math. Phys. 18, 1673 (1977) 10.1063/1.523468

\bibitem {sz3}S.W. Goode and J. Wainwright, Characterization of locally
rotationally symmetric space-times, Gen. Relativ. Gravit. 18, 315 (1986) 10.1007/BF00765890

\bibitem {sz5}J.A.S. Lima and J. Tiommo, Inhomogeneous two-fluid cosmologies,
Gen.\ Relat.\ Gravit. 20, 1019 (1988) 10.1007/BF00759023

\bibitem {rev1}K. Bolejko, M.-N. C\'{e}l\'{e}rier and A. Krasi\'{n}ski,
Inhomogeneous cosmological models: exact solutions and their applications,
Class. Quantum Grav. 28, 164002 (2011) 10.1088/0264-9381/28/16/164002

\bibitem {dd6}C. Saulder, S. Mieske, E. van Kampen and W.W. Zeilinger, Hubble
flow variations as a test for inhomogeneous cosmology, A\&A 622, A83 (2019) 10.1051/0004-6361/201629174

\bibitem {dd7}C. Clarkson and M.\ Regis, The cosmic microwave background in an
inhomogeneous universe, JCAP 02, 013 (2011) 10.1088/1475-7516/2011/02/013

\bibitem {dd8}K. Bolejko, The Szekeres Swiss Cheese model and the CMB
observations, Gen.\ Rel. Gravit. 41, 1737 (2009) 10.1007/s10714-008-0746-x

\bibitem {dd9}K. Bolejko and M. Korzy\'{n}ski, Inhomogeneous cosmology and
backreaction: Current status and future prospects, IJMPD 26, 1730011 (2017) 10.1142/S0218271817300117

\bibitem {per102}K. Bolejko and M.-N. C\'{e}l\'{e}rier, Szekeres Swiss-cheese
model and supernova observations, Phys. Rev. D 82, 103510 (2010) 10.1103/PhysRevD.82.103510

\bibitem {per103}M. Ishak and A. Peel, Growth of structure in the Szekeres
class-II inhomogeneous cosmological models and the matter-dominated era, Phys.
Rev. D 85, 083502 (2012) 10.1103/PhysRevD.85.083502

\bibitem {per104}D. Vrba and O. Svitek, Modelling inhomogeneity in Szekeres
spacetime, Gen. Relativ. Grav. 46, 1808 (2014) 10.1007/s10714-014-1808-x

\bibitem {as1}A. Gierzkiewicz and Z.A. Golda, On integrability of the Szekeres
system. I, Journal of Nonlinear Mathematical Physics 23, 494 (2016) 10.1080/14029251.2016.1237199

\bibitem {as2}A. Gierzkiewicz and Z.A. Golda, A complete set of integrals and
solutions to the Szekeres system, Phys. Lett. A 382, 2085 (2018) 10.1016/j.physleta.2018.05.038

\bibitem {as3}A. Paliathanasis and P.G.L. Leach, Symmetries and Singularities
of the Szekeres System, Phys. Lett. A 381, 1277 (2017) 10.1016/j.physleta.2017.02.009

\bibitem {ls0}A. Ramani, B. Dorizzi, B.\ Grammaticos and T. Bountis,
Integrability and the Painlev\'{e} property for low-dimensional systems, J.
Math. Phys. 25, 878 (1984) 10.1063/1.526240

\bibitem {as4}A. Paliathanasis, A. Zampeli, T. Christodoulakis and M.T.
Mustafa, Quantization of the Szekeres System, Class. Quantum Grav. 35, 125005
(20108) 10.1088/1361-6382/aac227

\bibitem {ls1}P.G.L Leach, Lie symmetries and Noether symmetries, Applicable
Analysis and Discrete Mathematics 6, 238-246 (2012) www.jstor.org/stable/43666171

\bibitem {as5}A.\ Paliathanasis, Quantum potentiality in Inhomogeneous
Cosmology, Universe 7, 52 (2021) 10.3390/universe7030052

\bibitem {as6}A. Zampeli and A. Paliathanasis, Quantization of inhomogeneous
spacetimes with cosmological constant term, Class. Quantum Grav. 38, 165012
(2021) 10.1088/1361-6382/ac1209

\bibitem {as7}J. Libre and C. Valls, On the dynamics of the Szekeres system,
Phys. Lett. A 383, 301 (2019) 10.1016/j.physleta.2018.10.050

\bibitem {as8}J. Libre and C. Valls, Dynamics of the Szekeres system, J. Math. Phys 62, 082502 (2021) 10.1063/5.0054051 

\bibitem {cop1}E.J. Copeland, A.R. Liddle and D. Wands, Exponential potentials
and cosmological scaling solutions, Phys. Rev. D 57, 4686 (1998) 10.1103/PhysRevD.57.4686

\bibitem {da1}A.A. Coley and R.J. van den Hoogen, The Dynamics of multiscalar
field cosmological models and assisted inflation, Phys. Rev. D 62, 023517 10.1103/PhysRevD.62.023517

\bibitem {da2}A.P. Billyard, A.A. Coley, J.E. Lidsey and U.S. Nilsson,
Dynamics of M theory cosmology, Phys. Rev. D 61, 043504 (2000) 10.1103/PhysRevD.61.043504

\bibitem {da3}A.A. Coley, Dynamical systems and cosmology, Astroph. Space Sci.
Libr. 291 (2003) 10.1007/978-94-017-0327-7

\bibitem {da4}G. Leon and E.N. Saridakis, Dynamical analysis of generalized
Galileon cosmology, JCAP 03, 025 (2013) 10.1088/1475-7516/2013/03/025

\bibitem {da5}G.A Rave-Franco, C. Escamilla-Rivera and J.L. Said, Dynamical
complexity of the teleparallel gravity cosmology, Phys. Rev. D 103, 084017
(2021) 10.1103/PhysRevD.103.084017

\bibitem {da6}R.G. Landim, Cosmological perturbations and dynamical analysis
for interacting quintessence, EPJC 79, 889 (2019) 10.1140/epjc/s10052-019-7418-8

\bibitem {da7}M. Kar\v{c}iauskas,\ Dynamical Analysis of Anisotropic
Inflation, Mod. Phys. Lett. A 31, 1640002 (2016) 10.1142/S0217732316400022

\bibitem {da8}G. Leon and E.N. Saridakis, Dynamical behavior in mimetic F(R)
gravity, JCAP 04, 031 (2015) 10.1088/1475-7516/2015/04/031

\bibitem {da9}A. Suroso and F.P.\ Zen, Cosmological model with nonminimal
derivative coupling of scalar fields in five dimensions, Gen. Rel. Gravit. 45,
799 (2013) 10.1007/s10714-013-1500-6

\bibitem {da10}A.\ Paliathanasis and G. Leon, Asymptotic behavior of N-fields
Chiral Cosmology, EPJC 80, 847 (2020) 10.1140/epjc/s10052-020-8423-7

\bibitem {da11}A. Giacomini, S. Jamal, G. Leon, A. Paliathanasis and J.
Saavedra, Phys. Rev.\ D 95, 124060 (2017) 10.1103/PhysRevD.95.064031

\bibitem{aulbach}
B. Aulbach, Continuous and Discrete Dynamics near Manifolds
of Equilibria (Lecture Notes in Mathematics No. 1058, Springer,
1984)

\end{thebibliography}
\end{document}